\documentclass[journal=dce]{CUP-JNL-DTM}

\usepackage{graphicx}
\usepackage{multicol,multirow}
\usepackage{amsmath,amssymb,amsfonts}
\usepackage{mathrsfs}
\usepackage{amsthm}
\usepackage{rotating}
\usepackage{appendix}
\usepackage{ifpdf}
\usepackage[T1]{fontenc}
\usepackage{newtxtext}
\usepackage{newtxmath}
\usepackage{textcomp}
\usepackage{xcolor}
\usepackage{lipsum}
\usepackage[colorlinks,allcolors=blue]{hyperref}
\usepackage{hyperref}
\usepackage{cleveref}
\usepackage{float}
\usepackage{array}

\theoremstyle{definition}

\numberwithin{equation}{section}

\jname{Data/Math}
\articletype{ARTICLE TYPE}
\jyear{2025}

\begin{document}

\begin{Frontmatter}

\title[Article Title]{Digital twin for virtual sensing of ferry quays via a Gaussian Process Latent Force Model}

\author[1]{Luigi Sibille}
\author[1]{Torodd Skjerve Nord}
\author[2]{Alice Cicirello}

\address[1]{Department of Ocean Operations and Civil Engineering, Norwegian University of Science and Technology, Ålesund, Norway}
\address[2]{Department of Engineering, University of Cambridge, Trumpington Street, Cambridge, UK}

\authormark{Luigi Sibille \textit{et al}.}
\authormark{Luigi Sibille et al.}


\abstract{Ferry quays experience rapid deterioration due to their exposure to harsh maritime environments and operational ferry impacts. Vibration-based structural health monitoring offers a valuable approach to assessing structural integrity and enhancing the understanding of the structural implications of these impacts. However, practical limitations often restrict sensor placement at critical locations. Consequently, virtual sensing techniques become essential for establishing a Digital Twin and estimating the structural response. This study investigates the application of the Gaussian Process Latent Force Model (GPLFM) for virtual sensing on the Magerholm ferry quay, combining in-operation experimental data collected during a ferry impact with a detailed physics-based model. The proposed Physics-Encoded Machine Learning model integrates a reduced-order structural model with a data-driven GPLFM representing the unknown impact forces via their modal contributions. Significant challenges are addressed for the development of the Digital Twin of the ferry quay, including unknown impact characteristics (location, direction, intensity), time-varying boundary conditions, and sparse sensor configurations. Results show that the GPLFM provides accurate acceleration response estimates at most locations, even under simplifying modeling assumptions such as linear time-invariant behavior during the impact phase. Lower accuracy was observed at locations in the impact zone. A numerical study was conducted to explore an optimal real-world sensor placement strategy using a Backward Sequential Sensor Placement approach. Sensitivity analyses were conducted to examine the influence of sensor types, sampling frequencies, and incorrectly assumed damping ratios. The results suggest that the GP latent forces can help accommodate modeling and measurement uncertainties, maintaining acceptable estimation accuracy across scenarios. 
}
\end{Frontmatter}

\section*{Impact Statement}
Ferry quays operate under severe environmental and operational conditions that often lead to deterioration before their intended design life. Accurate measurement of structural response is essential to improve understanding of their dynamic behavior and inform maintenance strategies. However, physical constraints, such as water exposure, vehicle traffic, and ferry impacts, prevent sensor placement at many critical locations. This study addresses these limitations by developing a Digital Twin framework for virtual sensing, integrating physics-based modeling with data-driven methods. The work highlights the modeling challenges specific to ferry quay infrastructure and provides a practical approach to estimate structural responses where instrumentation is not feasible. The findings support the development of more effective structural health monitoring practices for engineers, researchers, and infrastructure operators.

\section{Introduction}
Ferry quays play an essential role in the transportation networks of coastal and island regions worldwide. This infrastructure facilitates the transport of people, vehicles, and goods where the construction of traditional structures, such as bridges and tunnels, is not feasible. Particularly in Norway, these quays are critical, supporting over 130 routes and the transportation of approximately 40 million people annually. This usage is expected to increase with the growth in population, vehicle ownership, and tourism.

Operational loads, particularly the impacts exerted by ferries during docking, are acknowledged as significant contributors to the rapid degradation of these quays \citep{siedziako2023experimental}. These impacts may lead to brittle material failures, the initiation of fatigue cracks, buckling of elements, and the failure of joints, all of which compromise the reliability and integrity of the infrastructure \citep{siedziako2023experimental}. The structural responses during such events are not fully understood, despite their crucial importance for applications such as fatigue analysis, system identification, damage detection, and predictive maintenance, which are essential for extending the service life of these structures. Environmental factors, including saltwater corrosion, strong waves, and wind, exacerbate these effects, accelerating the degradation process and potentially reducing the lifespan of these structures \citep{XIA2019106946, WALL201321}. A notable incident exemplifying these vulnerabilities occurred during the storm Ingunn in 2024, which led to a failure of the linkspan at the Magerholm ferry quay shortly after the ferry undocked. The specific causes of this failure are yet to be determined, highlighting the importance of investigating how operational and environmental factors impact ferry quay safety. 

Ferry quays exhibit significant time-varying behavior driven by operational variability, such as interactions with ferries, vehicles, and environmental factors. The impact's intensities and characteristics for over 1500 impacts were recently assessed by means of temporal moments \citep{sibille2024measurement}. Although this approach captured the spatial and temporal characteristics of impact events, it circumvented the challenge of modeling the system's dynamic behavior. Environmental conditions, particularly tidal levels, have been demonstrated to influence the modal properties, highlighting the time-variant nature of these structures \citep{sibille2025IMAC}. Nonetheless, these studies do not address the challenges of modeling the complex, time-dependent system under real impact events, nor do they estimate structural responses at unmeasured or inaccessible locations.

Digital Twins (DTs) of dynamical systems have gained particular attention in recent years for understanding and predicting the behavior of complex engineering systems and critical infrastructure in operation \citep{ye2019digital}. A DT is a time-evolving digital model based on a two-way interaction between the physical system (usually monitored) and the digital counterpart, that can account for uncertainty, different systems interactions, and can be used for policy and decision making \citep{tao2022digital}. Consequently, the development of a DT based on integrating domain and expert knowledge with physics-based models, and with real-world data (often referred to as observations) obtained through Structural Health Monitoring (SHM), has emerged as a critical tool for guiding high-consequence decision making on complex engineering systems and critical infrastructure  \citep{haywood2024discussing}. Ferry quays exhibit complex and time-variant behavior, making the development of a DT particularly beneficial, as it provides engineers and decision-makers with an up-to-date understanding of the structure's actual condition and dynamic response. Unlike traditional Finite Element (FE) models, which do not account for temporal variability, model uncertainty, or incomplete knowledge of real boundary conditions, DTs provide a model that evolves in parallel with the physical system, incorporating real-world data from sensors and monitoring systems to continuously reflect the structure’s current state. To effectively build such a responsive and adaptive DT, particularly in environments with limited sensor coverage and complex loading conditions like ferry quays, Physics-Enhanced Machine Learning (PEML) strategies offer a powerful solution \citep{Cicirello_2024}. PEML strategies (also referred to as hybrid physics-data models, grey-box modeling, or scientific machine learning \citep{cross2024spectrum}) enable the development of robust and interpretable models based on the integration of physics-based models, knowledge, and real-world data (even if limited, sparse, and noisy). Traditional SHM systems, through sensor networks,  provide the measurement data that informs and continuously updates the DT \citep{Farrar2010}. However, the placement of sensors in areas subjected to impacts or traffic often presents practical challenges. Operational constraints and the risk of sensor damage in such environments frequently preclude their installation in critical locations. This scenario emphasizes the advantages of developing a DT for virtual sensing in SHM based on PEML strategies, which would provide reliable estimates of unmeasured structural responses such as strains, accelerations, and velocities. This DT would ideally leverage data from sensors located in accessible positions to estimate unobserved responses throughout the structure, including in critical yet hard-to-reach areas, offering a more comprehensive understanding of structural behavior while reducing costs and practical constraints associated with physical instrumentation. Moreover, a DT framework can be further leveraged to perform optimal sensor placement by identifying the most informative locations for instrumentation, thereby maximizing monitoring efficiency while minimizing cost and redundancy.

Virtual sensing methodologies can be broadly categorized into two approaches: deterministic and stochastic. Both usually rely on a physics-based model of the structure (hence “model-based” methods), but they differ in whether uncertainty is explicitly accounted for. Deterministic methods assume that the underlying system model is accurate and do not explicitly account for measurement noise and model errors. These methods typically employ direct mathematical relationships, such as pseudo-inverse matrices or mode shape interpolation, to reconstruct responses from measured data. Among the most widely used deterministic techniques is Modal Decomposition and Expansion (MD\&E), which was introduced by \citet{iliopoulos2016modal} for strain estimation in offshore wind turbine foundations. MD\&E operates by decomposing the structural response into a set of mode shapes, which are then expanded to estimate responses at unmeasured locations. While effective, applications of MD\&E to experimental case studies were found to be susceptible to low-frequency noise \citep{smyth2007multi}, particularly when estimating displacements from acceleration measurements. To address this, a multi-band MD\&E approach was later introduced \citep{iliopoulos2017fatigue}, which separately solved the inverse problem for different frequency bands. However, \citet{noppe2016full} identified discontinuities in the estimated responses at the frequency band boundaries, highlighting challenges associated with this method. Despite these limitations, MD\&E has been successfully applied to various structural configurations, including beams and plates \citep{rapp2009displacement, pelayo2015modal}, as well as monopile and tripod offshore platforms \citep{iliopoulos2016modal, tarpo2020expansion}. The primary advantages of deterministic methods such as MD\&E lie in their computational efficiency and ease of implementation, making them suitable for scenarios where the structural behavior remains well understood and stable over time. However, these approaches are highly sensitive to modeling inaccuracies and environmental variations, as they do not incorporate uncertainty in measurements or system dynamics. For this reason, deterministic methods are not well suited for the estimation of responses in operating ferry quays, where environmental and operational variations significantly influence the structural behavior.

Stochastic virtual sensing techniques treat unknown quantities, such as states, input forces, and sensor data, within a probabilistic framework. A key characteristic of stochastic methods is the reliance on recursive Bayesian estimation, where system states and external excitations are iteratively updated as new sensor data becomes available. Among these methodologies, Kalman filter-based \citep{kalman1960new} approaches have been extensively explored. One of the earliest implementations in structural dynamics was introduced by \citet{ma1998study}, who employed a Kalman filter coupled with a recursive least-squares estimator for force estimation problems. However, this method assumed the availability of displacement measurements at all degrees of freedom, which is often impractical in real-world applications due to the limited accessibility of measurement locations. This fundamental limitation motivated the development of the joint input-state estimator (JIS), also known as the Gillijns \& de Moor filter (GDF) \citep{gillijns2007unbiased, gillijns2007unbiased2}. This technique enabled the simultaneous estimation of inputs and states in linear systems while maintaining minimum variance and unbiasedness conditions. This algorithm was extended to the use of reduced-order models by \citet{lourens2012joint}. This enhancement enabled the use of a limited set of mode shapes, which reduced computational complexity. Additionally, the Augmented Kalman Filtering (AKF) approach was introduced to directly incorporate the inputs into the state vector \citep{lourens2012augmented}. However, this approach, like its predecessors, proved inadequate when applied to nonlinear or uncertain systems. This limitation was addressed in several extensions of the Kalman filter, including the Unscented Kalman Filter (UKF) \citep{chatzi2009unscented, wu2007application}, the Extended Kalman Filter (EKF) \citep{mariani2005impact}, and the particle filter \citep{chatzi2014nonlinear}. Although these methods proved effective in multiple applications, from lab experiments to large-scale implementations, they still exhibited challenges related to stability, sensitivity to noise, and the need for precise tuning of covariance parameters. Moreover, when applied to ill-conditioned systems or scenarios with limited or noisy measurement data, these methods can suffer from estimation drift or observability issues \citep{maes2015design, lourens2012augmented}. 

To address these challenges, necessary conditions for system inversion have been established to ensure instantaneous invertibility \citep{antsaklis1978stable, moylan1977stable}. These criteria include: (i) ensuring that dynamic forces and system states can be uniquely identified from the available sensor data, (ii) requiring that the inversion algorithm remains stable, so that minor perturbations in the measurements do not result in unbounded errors, and (iii) guaranteeing that the estimated states and forces are uniquely defined by the given measurements. These conditions were subsequently extended to reduced-order models in \citep{maes2015design}. However, practical implementations have demonstrated that controllability alone does not ensure accurate response estimation, particularly when environmental variations, measurement noise, or deviations from the assumed structural properties are present \citep{maes2016joint, maes2016verification, petersen2017estimation, aucejo2019practical}. This issue becomes particularly evident in applications where the location and direction of input forces are unknown, requiring the use of modal equivalent forces to approximate the unknown excitations \citep{lourens2019full}. One approach to partially relax these conditions involves allowing a time delay in estimation, as implemented through smoothers such as the Rauch-Tung-Striebel (RTS) smoother \citep{floquet2006state, maes2017validation}. Another approach, which has been shown to effectively address these challenges while preserving the advantages of simultaneous state and force estimation, is the use of Latent Force Models (LFMs) \citep{bilbao2022virtual}. 

LFMs establish a relationship between measured accelerations (the observed variables) and input forces and system states (the unobserved or latent variables) \citep{nayek2019gaussian}. The concept was first introduced by \citet{alvarez2009latent}, who demonstrated that treating the driving force in a second-order system as a Gaussian Process (GP) and updating it with observed responses significantly enhances performance compared to purely data-driven approaches. The unknown forces are represented as random Gaussian processes, with tunable covariance functions providing a non-parametric representation of the underlying physics governing the dynamic evolution of loads. While this approach allows for a more flexible and adaptive modeling of system inputs, early implementations suffered from high computational demands, as GP regression scales $O(N^3)$ with respect to the number of data points N. A significant development came from \citet{hartikainen2010kalman}, who showed that for certain covariance kernels, the GP can be converted into an equivalent state-space representation. This allowed GPLFMs to be implemented using Kalman filtering and smoothing algorithms. 

\citet{nayek2019gaussian} demonstrated that the GPFLM either outperforms or achieves comparable accuracy to conventional methods, such as the AKF, AFK-dm, and DKF, for joint input-state estimation across a range of loading scenarios. \citet{rogers2018bayesian, rogers2020application} explored their application for joint input-state-parameters estimation, while \citet{vettori2024assessment} analyzed the impact of different covariance functions on the performance of GPLFMs. In recent years, GPLFMs have been applied to several real-world use cases by developing hybrid physics-data models that combine physics-based models of complex systems with probabilistic machine learning descriptions of latent forces and measurements collected under operating conditions. In particular, GPLFMs have been applied to estimate and validate the strain responses in offshore wind turbines below the mud-line via virtual sensing \citep{bilbao2022virtual, zou2023virtual}, to characterize modal wind loads and modal states in long-span suspension bridges \citep{petersen2022wind}. These approaches have also been used in laboratory conditions, for evaluating the structural response of a 3D-printed scaled wind turbine blade \citep{vettori2024assessment}, and the discontinuous friction force of a single-storey frame with a brass-to-steel contact \citep{marino2023switching}.

In this paper, a DT of the Magerholm ferry quay is built based on the GPLFM and is presented for virtual sensing of acceleration responses in locations that are inaccessible or difficult to instrument due to operational constraints. To this end, \Cref{sec:math} presents the mathematical background for the GPLFM. \Cref{sec:case_study} introduces the Magerholm ferry quay case study, details the experimental data collected, outlines the challenges associated with virtual sensing, and discusses the modeling assumptions made. \Cref{sec:results} presents the results of the virtual sensing analysis, including an investigation into the influence of different sensor types, assumed damping ratios, and data sampling frequencies on the performance of the virtual sensing. Finally, \Cref{sec:conclusions} concludes the study.

\section{Summary of the building blocks for a ferry quay digital twin development for virtual sensing}\label{method_intro}
The development of the DT for ferry quays presented in this study is grounded in a hybrid modeling strategy that integrates physics-based structural dynamics with real-world data through PEML. This approach enables the construction of a dynamic digital of the ferry quay that evolves alongside the physical system under uncertain loading conditions and limited sensor availability.

The core of the DT is a GPLFM that leverages structural knowledge from a FE model of the structure while remaining adaptive to information obtained through monitoring data. Specifically, system matrices derived from the FE model represent the known physics of the structure, while latent forces, representing unknown external excitations such as impact loads, are modeled as GPs. This physics-encoded approach differs from physics-guided or purely data-driven methods in that the governing physics are directly embedded into the probabilistic model formulation, not just used to guide or regularize the learning. Consequently, it enables state and input estimation under sparse or noisy measurements while still maintaining physical interpretability. The Kalman filter and RTS smoother are used to infer both the system states and the latent forces, and the associated uncertainties are explicitly quantified through the process noise covariance $\mathbf{Q}$ and the measurement noise covariance $\mathbf{R}$. A key advantage of this DT framework is its efficiency with limited data. Unlike deep-learning-based DTs, which often require large volumes of training data to learn system behavior and may struggle when applied outside the conditions they were trained on, the GPLFM-based approach embeds structural knowledge directly into the model.

The DT is built through four key components:
\begin{itemize}
    \item \textbf{Physics-based structural model:} the structural characteristics are defined by system matrices derived from the FE model or operational modal analysis.
    \item \textbf{Data-driven latent force modeling:} unknown external forces are modeled as GPs whose characteristics are inferred from data. These GPs are then integrated into the overall state-space model via their own state-space representations.
    \item \textbf{Parameters tuning:} the GP kernel hyperparameters and the process noise covariance are tuned based on observed measurement data.
    \item \textbf{Response estimation:} the estimation of unmeasured responses (virtual sensing) and latent forces is achieved through the application of the Kalman filter and RTS smoother to the augmented state-space model, which updates predictions with the measurements.
\end{itemize}

Importantly, the framework accounts for multiple sources of uncertainty, including:
\begin{itemize}
    \item \textbf{Model uncertainty (e.g., discrepancies in FE model-based dynamics):} it is addressed through its associated process noise component.
    \item \textbf{Input uncertainty (e.g., variability in unmeasured impact forces):} it is captured by the GP model for latent forces and its associated process noise component.
    \item \textbf{Measurement noise (e.g., sensor inaccuracies):} it is represented by the measurement noise covariance matrix.
\end{itemize}

\section{Mathematical background of the Physics-Encoded Machine Learning strategy for virtual sensing}\label{sec:math}
\subsection{Mathematical model of structural system}
The equation of motion of a linear system with a $n_\text{dof}$ degree of freedom under forced vibration can be represented by a linear second-order ordinary differential equation as:
\begin{equation} \label{eq1}
    \mathbf{M}\ddot{\mathbf{u}}(t) + \mathbf{C}\dot{\mathbf{u}}(t) + \mathbf{K}\mathbf{u}(t) = \mathbf{S}_p \mathbf{p}(t),
\end{equation}
where $\mathbf{u}(t) \in \mathbb{R}^{n_\text{dof}}$ is the vector of displacements at $n_\text{dof}$ degrees of freedom, while the matrices $\mathbf{M}$, $\mathbf{C}$, $\mathbf{K}$ denote, respectively,  the mass, damping and stiffness of the system. The external forces applied to the system are represented by $\mathbf{S}_p \mathbf{p}(t)$ as a combination of loads. The Boolean input shape matrix $\mathbf{S}_p \in \mathbb{R}^{n_{\text{dof}} \times n_\text{p}}$ indicates the degree of freedom  where the $p^{th}$ load is applied, while the input vector $\mathbf{p}(t) \in \mathbb{R}^{n_\text{p}}$ represents the time history of the $p^{th}$ load.  

When it comes to complex systems, it is a common practice to represent the equation of motion by using a reduced order model as described in \citep{craig1985review, lourens2012joint}. The undamped eigenvalue problem associated with \cref{eq1} is expressed as: 
\begin{equation} \label{eq2}
    \mathbf{K} \boldsymbol{\Phi} = \mathbf{M} \boldsymbol{\Phi} \boldsymbol{\Omega}^2 
\end{equation}
which yields the eigenvector $\Phi_i$ as a column of the matrix $\boldsymbol{\Phi} \in \mathbb{R}^{n_{\text{dof}} \times n_{\text{dof}}}$ and the natural frequencies $\omega^2_i$ as diagonal of the matrix $\boldsymbol{\Omega}^2 \in \mathbb{R}^{n_{\text{dof}} \times n_{\text{dof}}}$. In the context of a reduced order model system, the dynamic behavior is captured by a reduced number of modes $n_r$ and the matrices are consequently truncated into $\boldsymbol{\bar{\Phi}} \in \mathbb{R}^{n_{\text{dof}} \times n_{\text{r}}}$ and $\boldsymbol{\bar{\Omega}}^2 \in \mathbb{R}^{n_{\text{r}} \times n_{\text{r}}}$. By premultiplying each term by $\boldsymbol{\bar{\Phi}^{\text{T}}}$ and transforming $\mathbf{u}(t) = \boldsymbol{\bar{\Phi}} \mathbf{r}(t)$, \cref{eq1} becomes: 
\begin{equation} \label{eq3}
    \boldsymbol{\bar{\Phi}}^{\text{T}}\mathbf{M}\boldsymbol{\bar{\Phi}}\ddot{\mathbf{r}}(t) + \boldsymbol{\bar{\Phi}}^{\text{T}}\mathbf{C}\boldsymbol{\bar{\Phi}}\dot{\mathbf{r}}(t) + \boldsymbol{\bar{\Phi}}^{\text{T}}\mathbf{K}\boldsymbol{\bar{\Phi}}\mathbf{r}(t) = \boldsymbol{\bar{\Phi}}^{\text{T}}\mathbf{S}_p \mathbf{p}(t).
\end{equation}
The equation of motion can be decoupled by leveraging the orthogonality conditions corresponding to a set of mass-normalized eigenvectors, such that $\boldsymbol{\bar{\Phi}}^{\text{T}}\mathbf{M}\boldsymbol{\bar{\Phi}}=\mathbf{I}$ and $\boldsymbol{\bar{\Phi}}^{\text{T}}\mathbf{K}\boldsymbol{\bar{\Phi}}=\boldsymbol{\bar{\Omega}}^2$. When assuming proportional damping, the modal damping matrix \(\boldsymbol{\bar{\Phi}}^\mathrm{T} \mathbf{C} \boldsymbol{\bar{\Phi}} = \boldsymbol{\bar{\Gamma}}\) 
is also diagonal. The matrix \(\boldsymbol{\bar{\Gamma}}\) includes the terms \(2 \xi_j \omega_j\) for \(j = 1, \dots, n_r\), where \(\xi_j\) is the modal damping ratio, \(\omega_j\) is the natural frequency, and \(n_r\) is the number of retained modes.

Consequently, the governing equation of motion in modal coordinates can be written as:
\begin{equation} \label{eq4}
    \ddot{\mathbf{r}}(t) + \boldsymbol{\bar{\Gamma}}\dot{\mathbf{r}}(t) + \boldsymbol{\bar{\Omega}}^2\mathbf{r}(t) = \boldsymbol{\bar{\Phi}}^{\text{T}}\mathbf{S}_p \mathbf{p}(t).
\end{equation}

Considering the forces transformed into their modal contributions by $\mathbf{f(t)}=\boldsymbol{\bar{\Phi}}^{\text{T}}\mathbf{S_p}\mathbf{p(t)}$, where $\mathbf{f(t)} \in \mathbb{R}^{n_{\text{r}}}$ and defining the state vector $\mathbf{x(t)} \in \mathbb{R}^{n_{\text{s}} \times n_{\text{s}}}$ as 
\begin{equation} \label{eq5}
    \mathbf{x}(t) = 
    \begin{bmatrix}
    \mathbf{r}(t) \\
    \dot{\mathbf{r}}(t)
    \end{bmatrix}    
\end{equation}
where $n_s=2n_{\text{r}}$, the continuous-time second-order differential equation, detailed in \cref{eq4}, can be written as a set of first-order continuous-time differential equations. These are so-called continuous-time state-space equations and are expressed as: 
\begin{equation} \label{eq6}
    \dot{\mathbf{x}}(t) = \mathbf{A}_c\mathbf{x}(t)+ \mathbf{B}_c\mathbf{f}(t),
\end{equation}
where the system matrices $\mathbf{A_c} \in \mathbb{R}^{n_s \times n_s}$ and $\mathbf{B_c} \in \mathbb{R}^{n_s \times n_r}$ are defined as: 
\begin{equation} \label{eq7}
    \mathbf{A}_c = 
    \begin{bmatrix}
    \mathbf{0} & \mathbf{I} \\
    -\boldsymbol{\bar{\Omega}}^{2} & -\boldsymbol{\bar{\Gamma}}
    \end{bmatrix}, \quad
    \mathbf{B}_c = 
    \begin{bmatrix}
    \mathbf{0} \\
    \mathbf{I}
    \end{bmatrix},
\end{equation}

When it comes to SHM systems, it's necessary to define a measurement model to describe the relationship between the state vector and the observed variables. These observed variables can be a linear combination of displacement, velocity, and acceleration and be arranged in a vector $\mathbf{y}(t) \in \mathbb{R}^{n_y}$ where $n_{\text{y}}$ represents the number of measurements. The data vector $\mathbf{y}(t)$ becomes:
\begin{equation} \label{eq8}
    \mathbf{y}(t) = \mathbf{S}_a \ddot{\mathbf{u}}(t) + \mathbf{S}_v \dot{\mathbf{u}}(t) + \mathbf{S}_d {\mathbf{u}}(t),
\end{equation}
where the selection matrices $\mathbf{S}_a \in \mathbb{R}^{n_a \times n_{\text{dof}}}, \mathbf{S}_v \in \mathbb{R}^{n_v \times n_{\text{dof}}}, \mathbf{S}_d \in \mathbb{R}^{n_d \times n_{\text{dof}}}$ are populated according to the location where acceleration, velocity and displacement (strains) are measured, respectively. 

The observation equation, expressed in \cref{eq8}, can be rewritten by utilizing the definition of the state vector $\mathbf{x}(t)$ as:
\begin{equation} \label{eq9}
    \mathbf{y}(t) = \mathbf{G}_c \mathbf{x}(t) + \mathbf{J}_c \mathbf{f}(t),
\end{equation}
where the output influence matrix \(\mathbf{G}_c\) and the direct transmission matrix \(\mathbf{J}_c\) are expressed as:
\begin{equation} \label{eq9}
    \mathbf{G_c} = 
    \begin{bmatrix}
    \mathbf{S}_d \boldsymbol{\bar{\Phi}}- \mathbf{S}_a \boldsymbol{\bar{\Phi}} \boldsymbol{\bar{\Omega}}^2  & \mathbf{S}_v \boldsymbol{\bar{\Phi}} - \mathbf{S}_a \boldsymbol{\bar{\Phi}} \boldsymbol{\bar{\Gamma}}
    \end{bmatrix}, \quad
    \mathbf{J_c} = 
    \begin{bmatrix}
    \mathbf{S}_a \boldsymbol{\bar{\Phi}}\boldsymbol{\bar{\Phi}}^{\text{T}}\mathbf{S_p}
    \end{bmatrix}.
\end{equation}
\Cref{eq6} and \cref{eq8} define the reduced order model of the continuous time-space representation of the system described in \cref{eq1}. However, measurement systems are not able to capture continuous-time observations $\mathbf{y}(t)$ in reality; instead, they sample the response of the system at discrete time intervals. Consequently, the continuous-time state space model is discretized in time intervals defined at $t_k = k{\Delta}t$ where $k = 1, \dots, N$ and expressed as:
\begin{equation} \label{structSSprocnoisematrices}
    \mathbf{x}_{k+1} = \mathbf{A} \mathbf{x}_k + \mathbf{B} \mathbf{f}_k + \mathbf{w_k}
\end{equation}
\begin{equation} \label{structSSmeasnoisematrices}
    \mathbf{y}_k = \mathbf{G} \mathbf{x}_k + \mathbf{J} \mathbf{f}_k + \mathbf{\eta_k}
\end{equation}
where $\mathbf{A} = e^{\mathbf{A}_c \Delta t}, \quad \mathbf{B} = [\mathbf{A} - \mathbf{I}] \mathbf{A}_c^{-1} \mathbf{B}_c, \quad \mathbf{G} = \mathbf{G}_c, \quad \mathbf{J} = \mathbf{J}_c.$ The random variables $\mathbf{w_k}$ and $\mathbf{\eta_k}$ account for process noise and measurement noise, respectively. These terms are assumed to be mutually correlated, zero-mean, white-noise signals with known covariances \(\mathbf{Q}^x = \mathbb{E} \{ \mathbf{w_k} \mathbf{w_k}^\top \}\) and \(\mathbf{R} = \mathbb{E} \{ \mathbf{\eta_k} \mathbf{\eta_k}^\top \}\).

\subsection{Gaussian process regression in state-space model}
GP regression is a non-parametric Bayesian regression method that provides a robust and probabilistic approach to modeling data \citep{williams1995gaussian, Rasmussen2004}. A GP is defined as a collection of random variables, any subset of which follows a joint Gaussian distribution. The GP is defined as: 

\begin{equation}
    f(t) \sim \mathcal{GP}(m(t), k(t, t',\theta)).
\end{equation}

where \(m(t)\) represents the expected value of the function, and \(k(t, t', \theta)\), also known as the kernel, defines the covariance between two points \(t\) and \(t'\) of the process. The properties of the GP, including smoothness, periodicity, and the scale of variation, are determined by the kernel. The hyperparameters \(\theta\) of the kernel control these properties, enabling the GP to adapt to the data's characteristics.

GPs with certain kernels that satisfy the Markov properties can be reformulated within a state-space model (SSM) framework \citep{hartikainen2010kalman, sarkka2013spatiotemporal}. This formulation replaces the conventional kernel-based representation with a dynamical system, where the GP is converted into a linear time-invariant stochastic differential equation (SDE) of order $\beta$: 
\begin{equation} 
\frac{d^\beta f(t)}{dt^\beta} + a_{\beta-1} \frac{d^{\beta-1} f(t)}{dt^{\beta-1}} + \cdots + a_0 f(t) = w(t), 
\end{equation}
where $f(t) \sim \mathcal{GP}(0, k(t, t'))$ is a GP with zero-mean and stationary covariance function $ k(t, t')$, $w(t)$ is zero-mean continuous-time white noise with spectral density \(S_w(w)=q_c\), and $\mathbf{a}(t) = \begin{bmatrix} a_{\beta-1}, a_{\beta-2}, \dots, a_{0} \end{bmatrix}^\top$ is a set of coefficients. The function \(f(t)\) and its derivatives \(\beta-1\) can be represented as a vector $\mathbf{z}(t) = \begin{bmatrix} f(t), \frac{df(t)}{dt}, \dots, \frac{d^{\beta-1}f(t)}{dt^{\beta-1}}\end{bmatrix}^\top$ and used to build a first-order vector Markov process as: 
\begin{equation}
\begin{aligned}\label{cGPSSM}
    \dot{\mathbf{z}}(t) = \mathbf{F}_c \mathbf{z}(t) + \mathbf{L}_c \mathbf{w}(t), \\
    \text{f}(t) = \mathbf{H}_c \mathbf{z}(t).
\end{aligned}
\end{equation}
where the matrices $\mathbf{F}_c \in \mathbb{R}^{n_\beta \times n_\beta}, \mathbf{L}_c \in \mathbb{R}^{n_\beta \times 1}, \mathbf{H}_c \in \mathbb{R}^{1 \times n_{\beta}}$ are given by,
\begin{equation}
    \mathbf{F}_c = \begin{bmatrix}
        0 & 1 & \cdots & 0 \\
        \vdots & \vdots & \ddots & \vdots \\
        0 & 0 & \cdots & 1 \\
        -a_0 & -a_1 & \cdots & -a_{\beta-1}
    \end{bmatrix}, \quad
    \mathbf{L}_c = \begin{bmatrix}
        0 \\
        \vdots \\
        0 \\
        1
    \end{bmatrix}, \quad
    \mathbf{H}_c = \begin{bmatrix}
        1 & \cdots & 0 & 1 
    \end{bmatrix}.
\end{equation}

Solving the SDE in \cref{cGPSSM} provides a complete probabilistic characterization of the state vector $\mathbf{z}(t)$, which is modeled as a Gaussian random variable with a mean $\hat{\mathbf{z}}(t)$ and covariance $\mathbf{P}$, such that $\mathbf{z}(t) \sim \mathcal{N}\left(\hat{\mathbf{z}}(t), \mathbf{P}\right)$.
For the specific case of linear time-invariant (LTI) SDEs with initial conditions $\mathbf{z}(t_0) \sim \mathcal{N}\left(\hat{\mathbf{z}}_0, \mathbf{P}_0\right)$, the mean and covariance of $\mathbf{z}(t)$ are defined by the following differential equations: 

\begin{subequations}
\begin{align}
\frac{d\hat{\mathbf{z}}(t)}{dt} &= \mathbf{F}_c \hat{\mathbf{z}}(t), \label{SDEmean} \\
\frac{d\mathbf{P}(t)}{dt} &= \mathbf{F}_c \mathbf{P}(t) + \mathbf{P}(t) \mathbf{F}_c^\text{T} + \mathbf{Q}_c. \label{SDEcov}
\end{align}
\end{subequations}

where the term $\mathbf{Q}_c=\mathbf{L}_c q_w \mathbf{L}_c^\text{T}$ represents the process noise covariance, which quantifies the uncertainty introduced by the white noise $w(t)$ in the system. 

For a zero-mean stationary process, the solution of the SDE reaches a steady-state form, where the state vector follows a constant Gaussian distribution, $\mathbf{z}_k \sim \mathcal{N}\left(\mathbf{0}, \mathbf{P}_\infty\right)$. Under these conditions, \cref{SDEcov} takes the form of the Lyapunov equation:
\begin{equation} \label{Lyapunov}
    0 = \mathbf{F}_c \mathbf{P}_\infty + \mathbf{P}_\infty \mathbf{F}_c^\text{T} + \mathbf{Q}_c
\end{equation}
The solution of \cref{Lyapunov}, which is a special case of Riccati equations, is usually computed using numerical algorithms to obtain the steady state covariance matrix $\mathbf{P}_\infty$ \citep{sarkka2019applied, grewal2014kalman}.

In order to use the Bayes' filter and perform GP regression, \cref{cGPSSM} needs to be discretized. By leveraging the Markov property, data is processed sequentially, allowing the posterior distribution of the state vector $\mathbf{z}_k$ to be estimated at each time step $k=0, \Delta t, \dots, T$. The discretized model is defined as: 
\begin{subequations}\label{dGPSSM}
\begin{align}
\mathbf{z}_{k} &= \mathbf{F}\,\mathbf{z}_{k-1} + \mathbf{L}\,w_{k-1} \label{dGPSSM_a}\\
f_{k} &= \mathbf{H}\,\mathbf{z}_{k}\label{dGPSSM_b}
\end{align}
\end{subequations}
where $\mathbf{F} = \exp(\mathbf{F}_c \Delta t)$, $\mathbf{L} = (\mathbf{F} - \mathbf{I})\mathbf{F}^{-1}_c\mathbf{L}_c$ and $\mathbf{H}=\mathbf{H}_c$.

The state-space representation in \cref{dGPSSM} provides a structured framework for Gaussian Process regression within a dynamic system. Specifically, \cref{dGPSSM_a} defines the dynamics model, which governs the evolution of the latent states over time. On the other hand, \cref{dGPSSM_b} serves as the measurement model, relating samples $y$ to the states. \Cref{dGPSSM} enables the update of the filtered posterior distribution of state vector $\mathbf{z}_k$ at the time step $k$ based on the filtered posterior distribution computed at the previous time step $k-1$. This process is carried out using forward filtering methods, such as the Kalman filter \citep{kalman1960new}. However, while the Kalman filter is optimal in a forward-prediction sense, it does not account for future observations that may further refine the estimates \citep{sarkka2008unscented}. To address this limitation, a backward smoother, such as the RTS smoother, can be applied \citep{rauch1965maximum}. The RTS smoother computes the smoothing distribution of $\mathbf{z}$, using future measurements to enhance state estimates retrospectively. 

The application of filters and smoothers requires defining the covariance $\mathbf{Q}^f$ of the discretized zero-mean Guassian white noise $w_{k-1}$, which can be expressed as:
\begin{equation}\label{dQ}
\mathbf{Q}^f = \int_{t_k}^{t_{k+1}} \exp\left(\mathbf{F}_c (\Delta t - \tau)\right) \mathbf{Q}_c \exp\left(\mathbf{F}_c (\Delta t - \tau)\right)^\text{T} d\tau
\end{equation}
For zero-mean stationary process, numerical algorithms can be used to retrieve the process noise covariance $\mathbf{Q}^f$ by solving the Lyapunov equation given by: 
\begin{equation}\label{dQss}
\mathbf{Q}^f = \mathbf{P}_\infty - \mathbf{F} \mathbf{P}_\infty \mathbf{F}^\text{T}
\end{equation}

\subsection{Gaussian process latent force model}
The GPLFM provides a probabilistic framework for estimating latent (unknown) forces acting on dynamical systems. Traditional SSMs rely on observed inputs to describe system dynamics. However, in many real-world applications, these inputs are either unknown or exhibit high complexity, making direct measurement or modeling impractical. The GPLFM addresses this issue by modeling latent forces as GPs and integrating them into a state-space formulation \citep{alvarez2009latent}. This hybrid approach takes advantage of both the mechanical system dynamics and the stochastic nature of the latent forces to infer the evolution of both system states and unknown inputs over time.

To achieve this, the state vector of the mechanical system, $\mathbf{x(t)}$ in \cref{eq6}, is augmented with the input state vector $\mathbf{z}$, which is derived from the GP representation in \cref{cGPSSM}. This results in an augmented state-space model, formulated as:
\begin{equation}\label{augmentedSSM}
\begin{aligned}
\begin{bmatrix}
\mathbf{\dot{x}}(t) \\
\dot{z}^{(1)}(t) \\
\dot{z}^{(2)}(t) \\
\vdots \\
\dot{z}^{(n_r)}(t)
\end{bmatrix}
&=
\underset{\mathbf{F}_c^a}{\underbrace{
\begin{bmatrix}
\mathbf{A}_c & \mathbf{b}_1 \mathbf{H}^{(1)}_c & \mathbf{b}_2 \mathbf{H}^{(2)}_c & \cdots & \mathbf{b}_{n_r} \mathbf{H}_c^{(n_r)} \\
0 & \mathbf{F}^{(1)}_c & 0 & \cdots & 0 \\
0 & 0 & \mathbf{F}^{(2)}_c & \cdots & 0 \\
\vdots & \vdots & \vdots & \ddots & \vdots \\
0 & 0 & 0 & \cdots & \mathbf{F}_c^{(n_r)}
\end{bmatrix}}}
\begin{bmatrix}
\mathbf{x}(t) \\
z^{(1)}(t) \\
z^{(2)}(t) \\
\vdots \\
z^{(n_r)}(t)
\end{bmatrix}
+
\underset{\mathbf{w}^a(t)}{\underbrace{
\begin{bmatrix}
\mathbf{0} \\
\tilde{\mathbf{w}}^{(1)}(t) \\
\tilde{\mathbf{w}}^{(2)}(t) \\
\vdots \\
\tilde{\mathbf{w}}^{(n_r)}(t)
\end{bmatrix}}}, \\[10pt]
\mathbf{y}(t) &= 
\underset{\mathbf{H}_c^a}{\underbrace{
\begin{bmatrix}
\mathbf{G}_c & \mathbf{j}_1 \mathbf{H}_c^{(1)} & \mathbf{j}_2 \mathbf{H}_c^{(2)} & \cdots & \mathbf{j}_{n_r} \mathbf{H}_c^{(n_r)}
\end{bmatrix}}}
\begin{bmatrix}
\mathbf{x}(t) \\
z^{(1)}(t) \\
z^{(2)}(t) \\
\vdots \\
z^{(n_r)}(t)
\end{bmatrix}
+ \mathbf{v}(t).
\end{aligned}
\end{equation}

where $\mathbf{b}_1, \mathbf{b}_2, \ldots, \mathbf{b}_{n_r}$ are the columns of the structural matrix $\mathbf{B}_c$, $\mathbf{j}_1, \mathbf{j}_2, \ldots, \mathbf{j}_{n_r}$ are the columns of the structural matrix $\mathbf{J}_c$, and $\mathbf{v}_k$ represents the measurement noise modeled as white Gaussian noise with covariance $\mathbf{R}$. (\cref{structSSmeasnoisematrices}).. $\tilde{\mathbf{w}}^{(j)}(t)=\mathbf{L}_c^j\mathbf{w}^{(j)}(t)$ is the white noise vector associated with each GP-driven latent force and with a spectral density $\mathbf{Q}^{(j)}$.\\
In shorthand, \cref{augmentedSSM} may be formulated as:
\begin{equation}
    \dot{\mathbf{s}}(t) = \mathbf{F}_c^a \mathbf{s}(t) +  \mathbf{w}_a(t)
\end{equation}
\begin{equation}
    \mathbf{y}(t) = \mathbf{H}_c^a \mathbf{s}(t) + \mathbf{v}(t)
\end{equation}
where the vector $\mathbf{s}(t) = \begin{bmatrix} \mathbf{x}(t), \mathbf{z}(t) \end{bmatrix}^\top \in \mathbb{R}^{2n_\text{s} \times 2n_\text{s}}$ incorporates both the states of the structural system and of the GP-driven latent forces, $\mathbf{F}_c^a \in \mathbb{R}^{2n_\text{s} \times 2n_\text{s}}$, and $\mathbf{H}_c^a \in \mathbb{R}^{n_y \times 2n_\text{s}}$. This hybrid physics-data model representation belongs to the so-called PEML approaches, which combine a state-space physics-based model of complex systems with a GP regression state-space model of the latent forces and measurements collected under operating conditions \citep{Cicirello_2024}.

The spectral density of the white noise vector, $\mathbf{w}_a(t)$, denoted as $\mathbf{Q}^a$, is formed by integrating the process noise covariance matrix $\mathbf{Q}^x$ of the structural system (\cref{structSSprocnoisematrices}) with the process noise matrix $\mathbf{Q}^f$ from the GP representation (\cref{dQss}). This results in:  
\begin{equation} \label{noice_matrices}
    \mathbf{Q}^{a} =
    \begin{bmatrix}
        {\mathbf{Q}}^x & 0 \\
        0 & {\mathbf{Q}}^f
    \end{bmatrix}
\end{equation}
To facilitate practical implementation, the continuous-time is discretized over finite time intervals where $\Delta t$ is the time discrete step. The augmented SSM in discrete form can be expressed as:
\begin{equation}
    \mathbf{s}_k^{a} = \mathbf{F}^{a} \mathbf{s}_{k-1}^{a} + {\mathbf{w}}_{k-1}^{a}
\end{equation}
\begin{equation}\label{augmentedmeas}
    \mathbf{y}_k = \mathbf{H}^{a} \mathbf{s}_k^{a} + \mathbf{v}_k.
\end{equation}
Once the system is discretized, its posterior distribution can be estimated using the Kalman filter at each time step, based on the current and past measurements. To further refine the state and latent force estimates, a backward smoothing step is applied using the RTS smoother.

\subsection{Choice of the covariance function and hyperparameters inference}\label{hyper}
An important step in the use of GPs involves selecting the covariance function or kernel. The kernel determines how values at different points relate to each other, influencing the smoothness, periodicity, and behavior of the GPs. To build an SSM of a GP, specific conditions regarding the kernel must be satisfied: the kernel must be semi-definite and should enable the GP to have a spectral density $S(w)$ in rational form. Once these criteria are met, the state-space matrices ($\mathbf{F}, \mathbf{L}, \text{and } q_w$) can be derived through spectral factorisation, as described in \citep{hartikainen2010kalman}.

In structural engineering applications, isotropic covariance functions, such as the periodic, squared exponential, exponential, and Matérn class, are usually employed \citep{vettori2024assessment}. The Matérn family of covariance functions is often the preferred choice in situations where there is limited prior knowledge about the loading conditions due to their versatility in modeling different types of random processes. \citet{vettori2024assessment} explored the inner dynamic features of the most common covariance functions and validated their performance on an experimental case study involving a 3D-printed scaled titanium WT blade under several loading conditions. Vettori et al. show that when the system is subjected to an impulse, or non-smooth load condition, the exponential function tends to perform best due to its ability to capture sharp variations. However, \citet{stein2012interpolation} argues that exponential kernels may impose unrealistic smoothness assumptions when applied to measured data from operating structures, potentially leading to inaccurate estimations. As an alternative, the Matérn 3/2 or 5/2 is often better suited. In this work, given the limited knowledge about the excitation, the Matérn 3/2 kernel is adopted, and its state-space representation is provided in \ref{app:ssm}.  

Once the kernel is chosen and its SSM representation is formulated, attention shifts to inferring its hyperparameters, which directly impact the quality of the GP model and the resulting GPLFM estimates. In the case of the Matérn 3/2 kernel, the hyperparameters govern the scale, length scale, and smoothness of the latent force estimations. A common approach is to tune them by using dynamic response measurements of the dynamical system. Ghahramani et al. \citet{ghahramani1996parameter} proposed a methodology to select the optimal hyperparameters maximizing the marginal likelihood of the data and improving the model's fit to the data. This method was integrated into a state-space model for dynamic systems by \citet{nayek2019gaussian}. The method consisted of minimizing the residual norm between observed data and model predictions based on the least squares solution. However, it has been observed to be susceptible to false estimations and not able to find the global maximum \citep{petersen2022wind}. To address this, various robust methodologies have been proposed over the years. \citet{rogers2020application} adopted a Bayesian framework with Monte Carlo simulations to obtain the full posterior distribution. Despite the methodology guarantees convergence to the true posterior distribution of interest, it is computationally expensive \citep{zou2023virtual, marino2023switching}. To reduce the computational cost of a switching GPLFM used to identify discontinuous forces in dynamical system, \citet{marino2023switching} introduced an approach that combines variational inference with active sampling Bayesian quadrature and used it in the context of mechanical systems with discontinuous linearities, the so-called Variational Bayes Monte Carlo \citep{acerbi2018variational, acerbi2020variational}. 

\citet{zou2023virtual} proposed a low-complexity and effective approach that is utilised in this study. This technique takes advantage of prior assumptions encoded in the GP kernel to align the empirical distribution of measurements, denoted as $ l = p(y) \sim \mathcal{N}(\mathbf{0}, \mathbf{P}^y)$, with the theoretical probability distribution of the Gaussian prior, defined as $ l^\theta = p(y|\theta) \sim \mathcal{N}(\mathbf{0}, \hat{\mathbf{P}}^y_\mathbf{\infty})$. Specifically, the hyperparameter inference is achieved by minimising the Hellinger distance between these two distributions, conducted as follows:
\begin{equation} \theta^* = \arg\min_\theta D(l, l^\theta), \end{equation}
where $D(l, l^\theta)$ represents the Hellinger distance. The empirical covariance of the measurements $\mathbf{P}^y$ is computed directly from the time series of sensor observations, while the covariance of the modeled Gaussian prior on the observed states $\hat{\mathbf{P}}^y_\mathbf{\infty}$, is obtained from the steady-state covariance of the augmented model, $\mathbf{P}_\mathbf{\infty}$,  which is encoded by the GP kernel function in \cref{Lyapunov}. This transformation is carried out using the matrix $\mathbf{H}^a$ of \cref{augmentedmeas} as:
\begin{equation} \hat{\mathbf{P}}^y_\mathbf{\infty} = \mathbf{H}^a \mathbf{P}_\mathbf{\infty} (\mathbf{H}^a)^\text{T}. \end{equation}

\subsection{Inference of the noise covariance matrices}
The process noise covariance matrix $\mathbf{Q}^x$ represents the uncertainty in the assumed dynamical model of the structure, primarily due to modeling errors. Lower (or higher) values of $\mathbf{Q}^x$ indicate more (or less) confidence in the dynamic model. In contrast, the input process noise covariance matrix $\mathbf{Q}^f$ represents the covariance of a hypothetical noise process that governs the evolution of the input estimates. The optimal tuning of these covariance matrices is a well-known challenge in algorithms such as AKF, AKF-dm, and DKF, considering their influence on the accuracy of the results.

In this study, the process noise covariance matrix $\mathbf{Q}$ was tuned using the same methodology employed for the GP kernel hyperparameters. Specifically, $\mathbf{Q}$ was optimised by minimising the Hellinger distance between the covariance of the Gaussian prior on the observed states and the empirical distribution of the measurements. In contrast, the measurement noise covariance matrix $\mathbf{R}$ was tuned by minimizing the RMSE between the posterior state estimates, obtained after Kalman filtering and RTS smoothing, and the measurements. Measurement noise tuning thus becomes an optimization problem formulated as:
\begin{equation}
R^* = \arg\min_R \text{RMSE}(\mathbf{y}^m, \mathbf{y}),
\label{eq:R_tuning_MSE}
\end{equation}
where $\mathbf{y}^m$ denotes the full sequence of measured outputs, and $\mathbf{y}$ is the sequence of predicted measurements obtained from the posterior estimates of the augmented state. At each time step $k$, the predicted measurement is given by $\mathbf{y}_k = \mathbf{H}^{a}\mathbf{s}_k^{a}$+$\mathbf{v}_k$, where $\mathbf{s}_k^{a}$ is the posterior estimate of the augmented state and $\mathbf{H}^{a}$ is the corresponding observation matrix. The measurement noise $\mathbf{v}_k$ is modeled as white Gaussian noise with zero mean and covariance $\mathbf{R}$ (as defined in \cref{structSSmeasnoisematrices}). The optimisation in \cref{eq:R_tuning_MSE} effectively tunes $\mathbf{R}$ by assessing the residuals between the measured data $(\mathbf{y}^m$ and the prediction of the observed variables $\mathbf{y})$.

\section{The Magerholm research quay case study: real-world data, FE model and challenges}\label{sec:case_study}
\subsection{Description of the Magerholm ferry quay}\label{num_descr}
The Magerholm ferry quay became a benchmark facility for SHM research through a collaboration between Møre og Romsdal county and NTNU in 2023 \citep{sibille2024measurement}. The quay ranks as one of the busiest ferry quays in Norway, transporting approximately 950,000 vehicles across the Storfjorden fjord in 2023. Two ferries operate on this route, each measuring 111 meters in length, with a gross tonnage of 3,000, and capable of transporting up to 120 cars. The operational schedule varies throughout the week, with 55 crossings per day on weekdays and 48 crossings on weekends. The frequency of crossings changes throughout the day, with the longest break occurring at night (approximately two hours). During peak hours, departures take place every 20 minutes.

\begin{figure}
    \centering
    \includegraphics[width=1\linewidth]{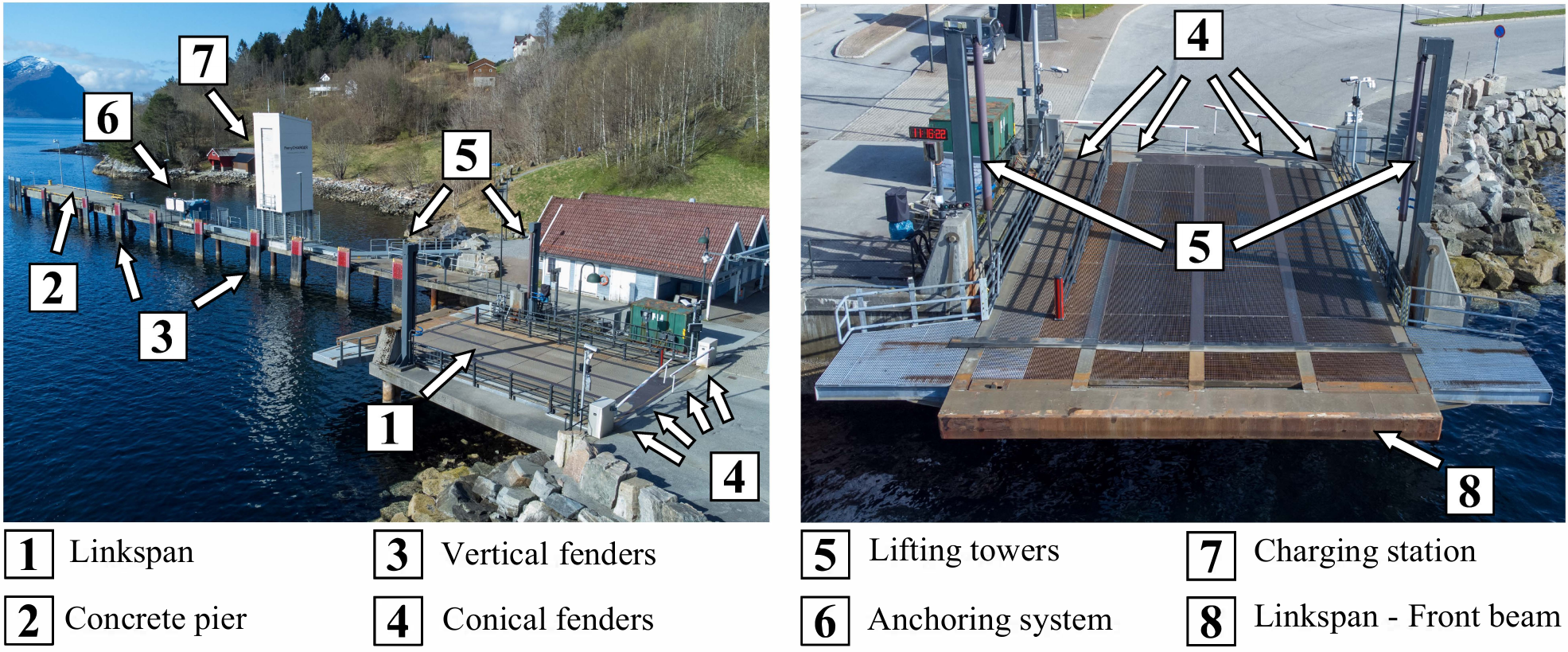}
    \caption{The Magerholm ferry quay}
    \label{fig_magerholm_quay}
\end{figure}

The quay, depicted in \Cref{fig_magerholm_quay}, consists of two independent structures. The first is a concrete pier approximately 110 meters long, supported by concrete-filled steel piles. It is fitted with thirteen fenders arranged to absorb the forces generated during ferry docking. The second structure is a steel linkspan designed for loading and unloading vehicles and passengers. The linkspan measures approximately 15 meters in length and 9 meters in width, composed of five HE550B longitudinal beams and IPE 240 cross beams. Due to its exposure to frequent ferry impacts and significant wear, the front beam is made from Hardox400 steel. The linkspan is vertically supported by a concrete abutment and two hydraulic lifting towers, which allow height adjustments to accommodate water level fluctuations. The longitudinal movements (surge direction) are absorbed by four conical fenders positioned between the steel structure and the concrete foundation. The quay also includes an electric charging station and a vacuum anchoring system designed to stabilize the ferry when docked, reducing its movements. For a comprehensive description of the quay, readers can refer to \citet{sibille2024measurement}.

\subsection{Measurement campaign and data processing}
This study focuses exclusively on the linkspan, while the general layout of the Magerholm ferry quay is outlined for context. The linkspan was selected due to its higher susceptibility to structural degradation, more demanding maintenance routines, and notably greater dynamic responses compared to the concrete pier.

\begin{figure} [H]
    \centering
    \includegraphics[width=0.8\linewidth]{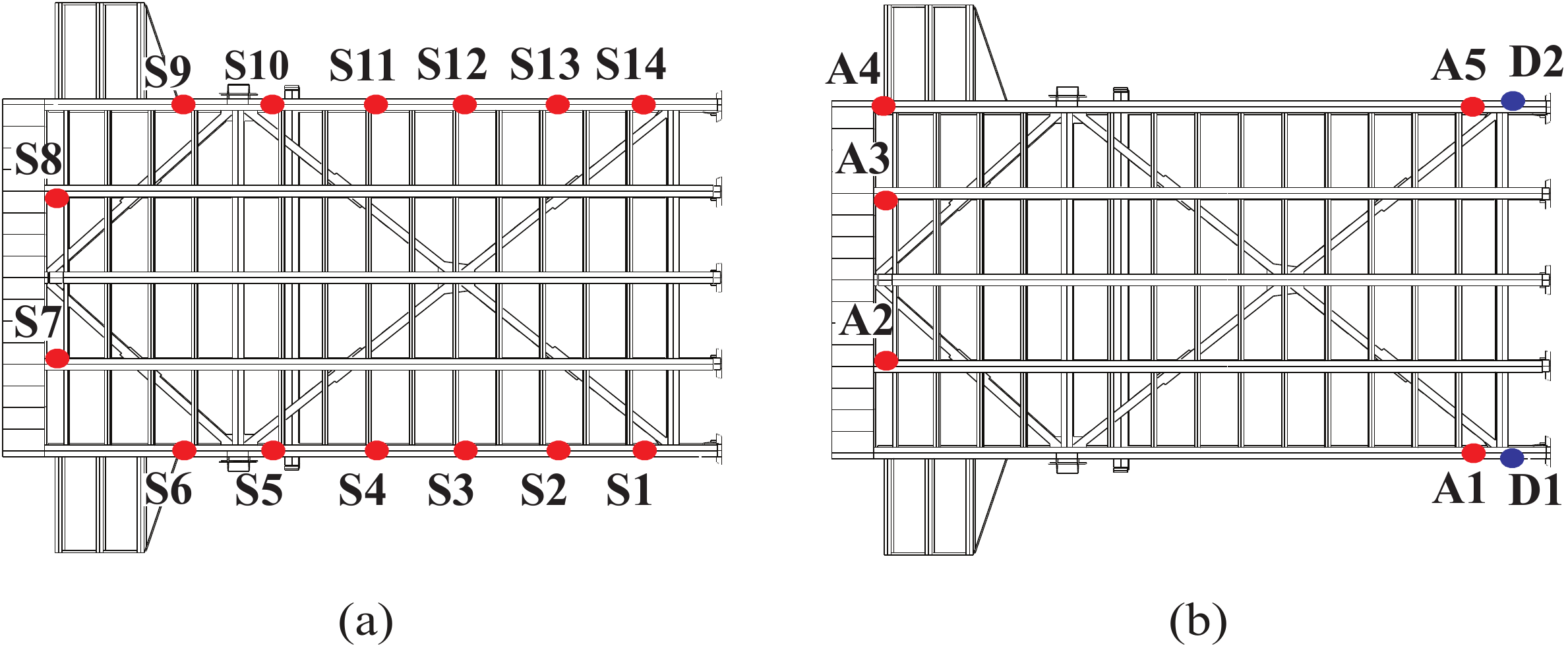}
    \caption{Sensor positions on the linkspan: (a) sensor configuration 1; (b) sensor configuration 2.}
    \label{fig_sens_pos}
\end{figure}

Response data were collected on the linkspan during two distinct ferry impact events, using two different data acquisition configurations shown in \Cref{fig_sens_pos}.a and \Cref{fig_sens_pos}.b. The instrumentation consisted of G-Link-200 3-axis accelerometers, housed in waterproof enclosures provided by Parker LORD MicroStrain, and Linear Variable Differential Transformers (LVDTs) supplied by Load Indicator System AB. The accelerometers' range was configured to ±2g with a noise density of $25~\mu g/\sqrt{\text{Hz}}$. The LVDTs had a nominal stroke of 100 mm. For both configurations, all sensors transmitted data to a laptop through a WSDA-2000 gateway via a wireless connection. To reduce low-frequency noise and potential drift, all recorded acceleration signals analyzed in \Cref{sec:results} underwent high-pass filtering with a cutoff frequency of 0.5 Hz. The displacement data were not filtered to preserve the quasi-static component of the linkspan's response.

The monitoring campaign was split into two distinct acquisition setups due to the data logger's capacity limitations, which restricted the employment of many sensors at high sampling frequencies. One setup focused on maximizing spatial information by using a larger number of accelerometers sampling at a moderate frequency (124 Hz), while the other prioritized temporal resolution by capturing acceleration and displacement responses at a higher sampling rate (1024 Hz). This approach enabled a comprehensive assessment of how sensor positions, types (accelerometers and LVDTs), and sampling frequencies influence the performance of virtual sensing. In both setups, accelerometer data were recorded only along the longitudinal axis (aligned with the direction of the expected impact) to maximize the number of operating sensors and sampling frequency within the data logger's constraints.

The two data acquisition setups were as follows:
\begin{itemize}
    \item  \textbf{Configuration 1 (Impact event 1):} this configuration involved fourteen G-Link-200 accelerometers, positioned as shown in Fig. 2.a. These sensors were set up to sample at 128 Hz.
    \item \textbf{Configuration 2 (Impact event 2):} this configuration included five G-Link-200 accelerometers and two LVDTs, with their locations depicted in Fig. 2.b. These sensors recorded data at a sampling frequency of 1024 Hz. The LVDTs were positioned to measure the relative movement of the linkspan with respect to the concrete abutment.
\end{itemize}

\begin{figure} [H]
    \centering
    \includegraphics[width=1\linewidth]{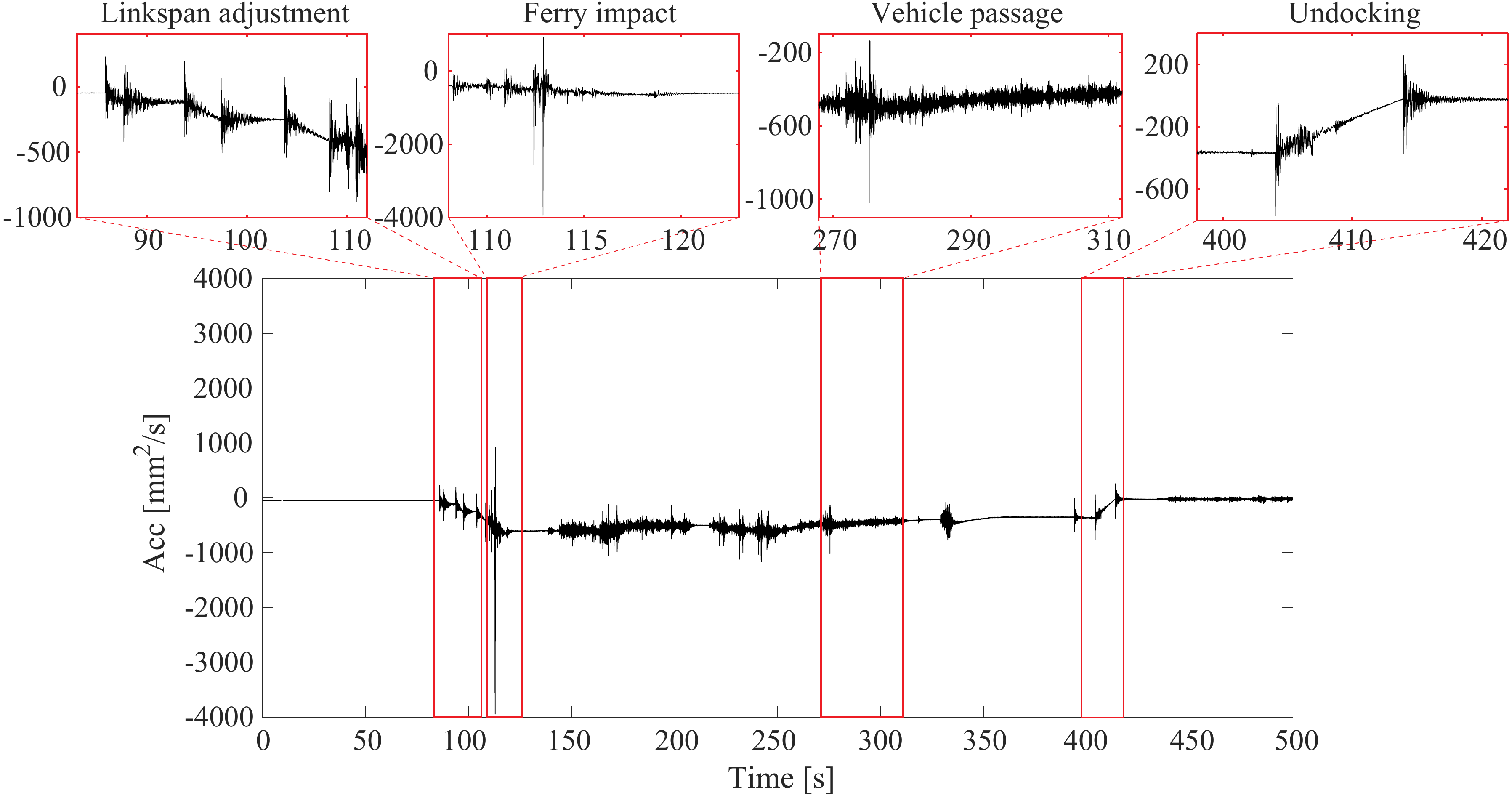}
    \caption{Full time series recorded from sensor S1 in configuration 1 during a complete docking event}
    \label{fig_docking}
\end{figure}

The complete docking can be categorized into four distinct phases: (1) adjustment of the linkspan prior to impact, (2) ferry impact, (3) vehicle passage, and (4) undocking. The unfiltered acceleration response recorded by sensor S1 in configuration 1 during these four phases is shown in  \Cref{fig_docking}. The adjustment phase occurs as the ferry approaches the quay. Because the clearance between the ferry and the quay varies with the tidal level, the linkspan must be manually repositioned to align with the ferry ramp. An operator onboard the ferry performs this adjustment to ensure that the impact occurs at a suitable location on the linkspan. This phase is reflected in the acceleration signals as changes in the mean value since the gravitational component is captured by the sensors. The second phase is generally short in duration, as it typically lasts only a few seconds. However, it results in the highest acceleration peaks due to the impact between the ferry and the structure. The third phase involves the unloading and loading of passengers and vehicles. This phase is significantly longer, lasting several minutes (approximately 300 seconds in this case). The acceleration response during this period is more moderate, though high accelerations can be observed when heavy vehicles, such as trucks, pass near the sensors. Finally, the ferry undocking phase marks the return of the linkspan to its initial, unloaded configuration. This phase is typically characterized by a gradual change in acceleration levels as the ferry undocks and the linkspan readjusts its position.

The distinct operational phases, particularly the impact and vehicle passage, introduce significant complexities for system modeling. During adjustment and undocking, the changing inclination of the linkspan occurs. While variations in the stiffness of structural elements like conical fenders or lifting towers due to these inclination change might be considered minor, the more dominant challenge arises from the interaction between the linkspan and the ferry. This interaction alters the linkspan's support configuration. This creates time-varying boundary conditions and couples the linkspan's dynamics with the ferry's movements, which are influenced by unpredictable environmental factors (e.g., waves, wind) and operational variables (e.g., engine thrust) \citep{sibille2024measurement, sibille2025IMAC}. Given these complexities, and to provide a focused evaluation of the GPLFM capabilities for virtual sensing under real conditions of unknown impact loading and inherent model uncertainties, this study concentrates its analysis exclusively on the ferry impact phase. For this specific time window, the structural system of the linkspan is assumed to behave as a Linear Time-Invariant (LTI) system. This simplification is adopted because precisely modeling the rapidly evolving contact mechanics during the impact event (e.g., determining whether it behaves as a perfect impact, quantifying a coefficient of restitution, or timing the exact moment of impact) presents considerable challenges. A detailed analysis of the physics of the impact is beyond the scope of this paper, which focuses on virtual sensing.

\subsection{FE model}
In this work, the system matrices $\mathbf{A}$ and $\mathbf{B}$ used in \cref{augmentedSSM}, representative of the system's dynamics and its relationship to external inputs, are retrieved from a FE model of the quay. The FE model is developed in Abaqus based on construction drawings and field measurements. The model includes only structural elements, such as HE550B longitudinal beams, IPE240, HE240A, and UNP220 secondary beams, all of which are modeled using S235 and S275 steel. The mesh is generated using quadrilateral finite membrane-strain shell elements. Additionally, the mass of non-structural elements, including railings and gratings, is incorporated, resulting in a total modeled structural weight of 53.64 tons. The conical fenders are modeled as longitudinal springs with a stiffness of 2000 KN/m. The lifting towers are represented as vertical springs with a stiffness set at 10000 KN/mm \citep{Siedzakomodal, leIMAC}. Considering that the four conical fenders are designed to engage sequentially based on displacement, only two of the four fenders were considered active in the FE analysis. For the impact events analyzed in this study, this assumption was justified, as the recorded displacements did not reach the threshold required to activate the remaining two fenders, allowing the system to be reasonably approximated as a linear system supported by two fenders. The model is updated using a sensitivity-based FE model updating method with modal parameters obtained from operational modal analysis \citep{leIMAC}. During the optimization process, the following parameters are defined as design variables: steel density, Young's modulus, and the stiffness of the two lifting towers.
 
\begin{figure} [H]
    \centering
    \includegraphics[width=0.85\linewidth]{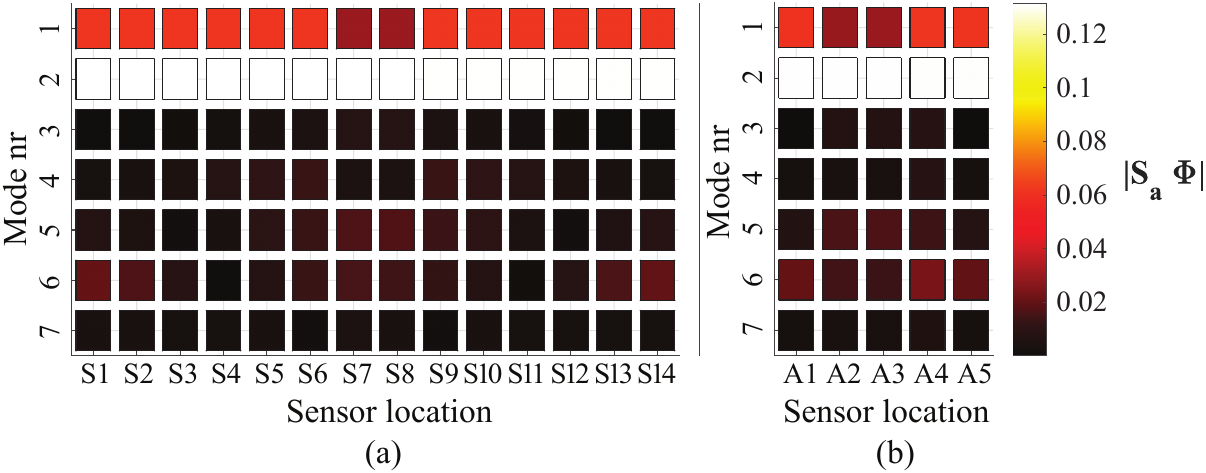}
    \caption{Modal influence of accelerations for: a)sensor configuration 1; b)sensor configuration 2}
    \label{fig_modalinfluence}
\end{figure}

To facilitate computational efficiency, a reduced-order model is constructed by retaining the first seven most influential modes identified from the FE Model. These modes, whose mode shapes are presented in the \ref{app:mode_shapes}, correspond to a frequency range extending up to 10.10 Hz. \Cref{fig_modalinfluence} illustrates the modal participation at each sensor location for these seven modes, specifically for sensor configuration 1 (Fig. \ref{fig_modalinfluence}a with fourteen sensors S1-S14) and sensor configuration 2 (Fig. \ref{fig_modalinfluence}b with five sensors A1-A5). The color intensity in the plots represents the magnitude of the product of the sensor selection matrix ($\mathbf{S_a}$) and the mode shape matrix ($\boldsymbol{\Phi}$), indicating the sensitivity of each sensor to each mode. The figure shows that the first two modes, which are predominantly horizontal, exhibit the highest influence across sensor locations in both configurations. This is consistent with the accelerometers recording primarily along the longitudinal axis. Among the other modes, which represent vertical/torsional behaviors, the third and seventh modes show the lowest influence on the sensors. 

Furthermore, due to the absence of experimental modal damping estimates in the horizontal direction,  a uniform modal damping ratio of 5\% is assumed for all modes included in the reduced order model. This damping ratio is used to build the modal damping matrix $ \boldsymbol{\bar{\Gamma}}$ in \cref{eq4}.

\subsection{Challenges to data and model assumptions}
The application of the GPLFM to the Magerholm ferry quay for virtual sensing faces various challenges. These arise from: (i) uncertainties in accurately representing the structural model, (ii) the unknown characteristics of the impact force, (iii) the time-variability of the system, especially during its interaction with the ferry, and (iv) spatially sparse measurements. The following points detail how each key challenge is accounted for:
\begin{itemize}
    \item \textbf{Modeling Uncertainties:} The structure incorporates numerous steel components connected through joints, which are subjected to a harsh maritime environment. This exposure causes corrosion, altering material properties and joint behavior in ways that are difficult to quantify precisely.  These uncertainties are partially mitigated by performing FE model updating using parameters derived from OMA data \citep{leIMAC}.
    \item \textbf{Unknown impact force characteristics:} The unknown magnitude, duration, and specific time history of the impact force are directly handled by modeling the force as a GP latent force within the GPLFM. The unknown location and direction of the impact force are addressed by decomposing the latent forces $\mathbf{f(t)}$ into its modal components. This modeling approach involves two critical assumptions. First, the modal components of the forces are assumed to be uncorrelated, which simplifies the estimation process but may not always be realistic when dealing with experimental data, as structural systems often exhibit coupling between dynamic modes. Second, all modal components are modeled using the same GP prior, characterized by identical kernels and hyperparameters. Although the same state-space matrices are employed for all the latent forces and calculated using the same kernel and hyperparameters, the posterior variance of the states of the modal components of the latent forces is computed via Kalman filter and smoothing. \citet{zou2023virtual} demonstrated that this assumption does not affect the estimation of the posterior estimates of the latent states and the performance of the GPFLM, as they have assigned different hyperparameters for each latent force and assessed the accuracy of the GPLFM.
    \item \textbf{System time-variability and boundary conditions:} The system experiences considerable time-variability due to the alteration of the support configuration. As illustrated in \Cref{fig_operational_states}, the linkspan switches from a 'disconnected state' (\Cref{fig_operational_states}a), where it is supported by the abutment (A) and the hydraulic cylinders (C), to a 'docked state' (\Cref{fig_operational_states}b), where the linkspan rests on the ferry. This shift in boundary conditions not only alters the physical supports but also dynamically couples the linkspan's behavior with the ferry’s movements, which are influenced by variable environmental conditions (e.g., wave action, wind) and operational inputs (e.g., engine thrust) \citep{sibille2024measurement, sibille2025IMAC}. These transitions alter the modal properties, such as natural frequencies, damping ratios, and mode shapes, of the linkspan within short timeframes. For the scope of this study, the system is assumed to be an LTI during the analyzed time window. While this is acknowledged as a strong assumption, it allows for avoiding introducing a switching system that would entail its own set of challenging unknowns, as the estimation of the modal properties for the different states within the impact and of the precise timing of these switches, which are beyond the primary focus of this study. Moreover, modal damping ratios are assumed quantities due to a lack of direct operational identification for all relevant modes/directions. Consequently, the sensitivity of the GPLFM’s virtual sensing performance to variations in the assumed damping ratio is evaluated to understand the potential impact of this uncertainty in \Cref{sec:dr_influence}.
    \begin{figure} [h]
        \centering    \includegraphics[width=1\linewidth]{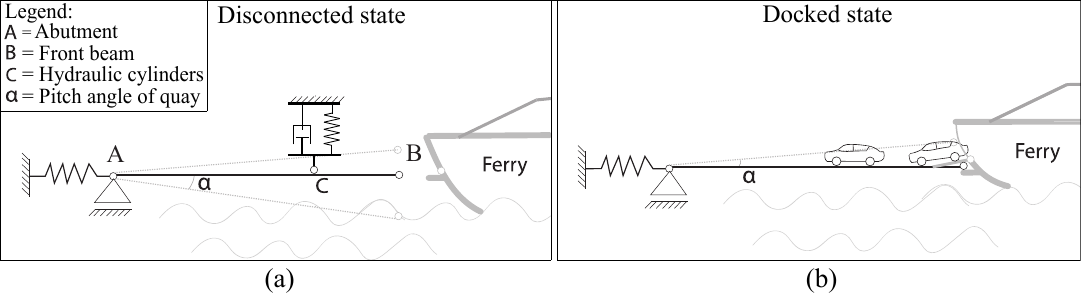}
        \caption{The operational states of the linkspan: a) Disconnected state; b) Docked state}
        \label{fig_operational_states}
    \end{figure}
    \item \textbf{Spatially sparse measurements and uni-axial data:} 
    The data acquisition setup comprises a limited number of sensors, leading to spatially sparse measurements across the structure. Moreover, the available sensors record acceleration data exclusively along the longitudinal direction (x-axis) of the linkspan. This uni-axial measurement constraint implies that dynamic responses in the transverse (y-axis) and vertical (z-axis) directions are not directly captured.
    \end{itemize}

These assumptions, along with the corresponding analyses performed in this study, are summarized in Table~\ref{tab:challenges_summary}.

\begin{table}[H]
\centering
\caption{Summary of key challenges: assumptions made and performed analyses}
\label{tab:challenges_summary}
\begin{tabular}{@{}>{\raggedright\arraybackslash}p{0.25\textwidth}>{\raggedright\arraybackslash}p{0.37\textwidth}>{\raggedright\arraybackslash}p{0.28\textwidth} @{}}
\toprule
\textbf{Challenge} & \textbf{Assumptions made} & \textbf{Analyses performed} \\
\midrule
\textbf{Modeling uncertainties} & 
Partial mitigation via FE Model Updating (OMA-based) & -- \\
\addlinespace
\textbf{Unknown impact force characteristics} & 
Force magnitude/history as GP latent force. \newline Force location/direction via modal decomposition of latent forces. & Investigation of GPLFM performance under varying assumed impact locations. \\
\addlinespace
\textbf{System time-variability and boundary conditions} & System assumed as LTI during the ferry impact. & Sensitivity analysis of GPLFM performance to variations in assumed damping ratios. \\
\addlinespace
\textbf{Spatially sparse measurements and uni-axial data} & 
Dynamic responses in the transverse (y-axis) and vertical (z-axis) directions are not directly included in the GPLFM. & Sensitivity analysis of GPLFM performance to sensor locations. \\
\bottomrule
\end{tabular}
\end{table}

\section{Virtual Sensing via GPLFM}\label{sec:results}
\subsection{Influence of sensor location on the accuracy of the estimated response}\label{sec:sensor_location_influence}
A critical aspect influencing the robustness of any virtual sensing methodology is the spatial relationship between the available instrumented sensors and the location where unmeasured responses are to be estimated. To explore this sensitivity within the context of the GPLFM, the accuracy of the estimated acceleration response is evaluated as it varies with the target sensor location. For this investigation, sensor configuration 1, comprising fourteen accelerometers (S1-S14) as illustrated in \Cref{fig_sens_pos}, was selected. 

The accuracy of the estimated acceleration response at each sensor location was evaluated using the RMSE. The evaluation was performed using a leave-one-out cross-validation approach: for each sensor location $S_i$ (where $i$ ranges from 1 to 14) within sensor configuration 1, the measured data from that specific sensor $S_i$ was temporarily excluded from the set of observations. The GPLFM was then performed using the measurement data from all other thirteen available sensors (i.e., sensors $S_1, \dots, S_{i-1}, S_{i+1}, \dots, S_{14}$) to perform the joint input-state estimation process. Based on these measurements, the GPLFM then estimated the acceleration response at the location of the excluded sensor $S_i$. This estimated response was compared in the time domain against the measured response from $S_i$ using the RMSE metric. This entire process was repeated iteratively for each of the fourteen sensor locations.

\begin{figure} [H]
    \centering    \includegraphics[width=0.8\linewidth]{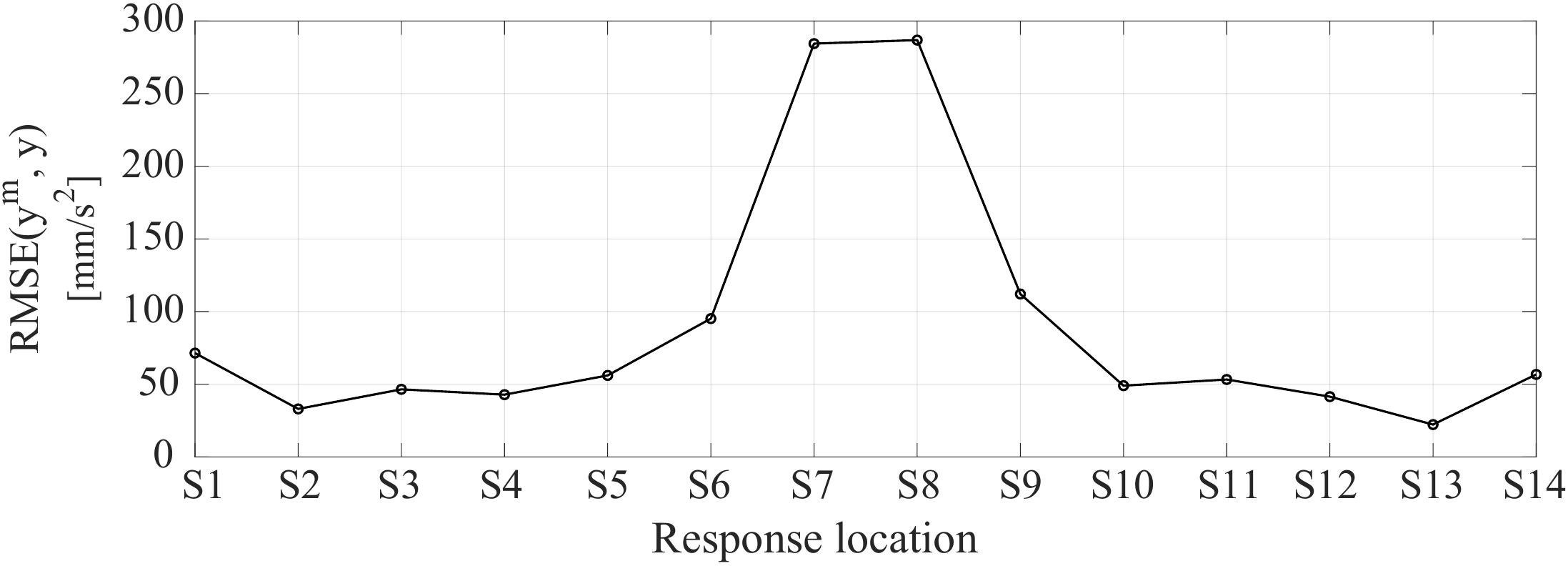}
    \caption{Root Mean Squared Error (RMSE) for acceleration response estimation at each sensor location (S1-S14) using a leave-one-out approach with sensor configuration 1.}
    \label{fig:rmse_leave_one_out}
\end{figure}

\Cref{fig:rmse_leave_one_out} shows the variation in estimation accuracy across the different sensor locations. Overall, the RMSE values are generally below 150 mm/s$^2$ for most locations. However, locations near the edges of the longitudinal beams, such as S1, S6, S9, and S14, tend to exhibit slightly higher RMSE values compared to sensors located more centrally along the beams (e.g., S2-S5, S10-S13). This may be attributed to the fact that these sensors have fewer neighboring sensors to provide spatial information. Moreover, a notable increase in RMSE is observed at locations S7 and S8, which are situated on the front beam of the linkspan.

\begin{figure} [H]
    \centering    \includegraphics[width=1\linewidth]{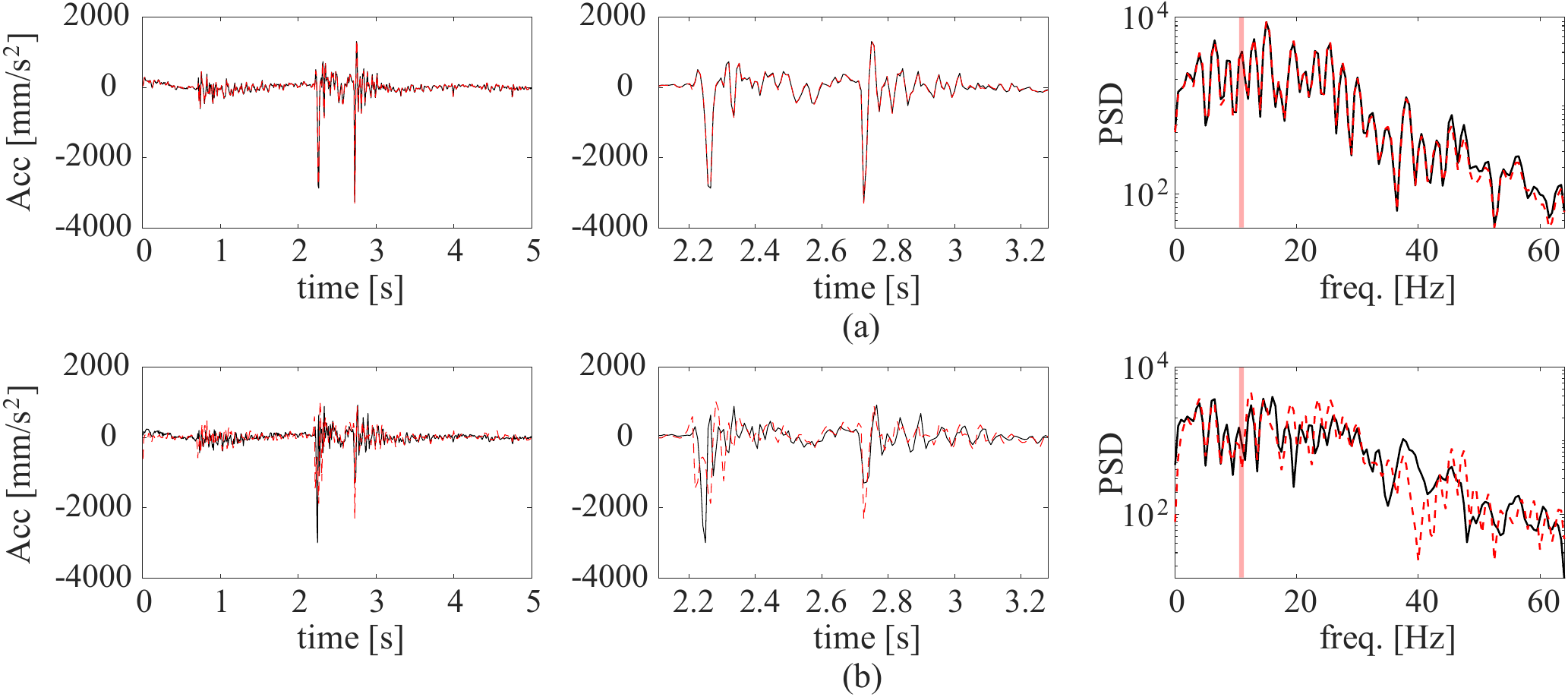}
    \caption{Full time series (left column), zoomed-in view of the time series (center column), and power spectral density (right column) for estimated
(black solid lines) and measured (dashed red line) acceleration responses: (a) at location S13; (b) at location S8. The red vertical bar in the PSD plots indicates the natural frequency of the highest mode included in the reduced-order model.}
    \label{fig:time_freq_S1_S8}
\end{figure}

To further illustrate these performance differences, \Cref{fig:time_freq_S1_S8} compares the estimated and measured acceleration responses in both the time and frequency domains for two representative cases. Specifically, \Cref{fig:time_freq_S1_S8}a shows the results for location S13 (corresponding to the lowest RMSE), while \Cref{fig:time_freq_S1_S8}b showcases the performance at location S8 (corresponding to the highest RMSE).

As shown in \Cref{fig:time_freq_S1_S8}.a (location S13), the GPLFM demonstrates high accuracy in reconstructing the acceleration response. The estimated time history closely matches the measured signal and both the amplitude and timing of the peak acceleration during the ferry impact are accurately captured. Furthermore, the Power Spectral Density (PSD) of the estimated response shows high agreement with that of the measured data across the entire analyzed frequency spectrum. In contrast, the estimation for location S8, shown in \Cref{fig:time_freq_S1_S8}b, presents inaccuracies, particularly in correspondence with the high acceleration peaks of the impact. However, despite the high RMSE, the timing of the acceleration peak of the ferry impact is reasonably well-identified, although its amplitude is not well reconstructed. In the frequency domain, the estimated PSD for S8 aligns reasonably with the measured PSD, below approximately 10 Hz. Beyond this frequency, the estimated PSD deviates, failing to capture the energy content at higher frequencies. While an initial inaccuracy is noticeable around 10 Hz, which roughly corresponds to the upper frequency limit of the reduced-order model used in this study (retaining modes up to approximately 10.10 Hz), the discrepancy becomes particularly pronounced after approximately 20 Hz. 

The observed lower accuracy in reconstructing the response at locations S7 and S8, both situated on the front beam of the linkspan may be attributed to: (i) ill-conditioning with distance from impact/observation locations; (ii) sensor distribution across different longitudinal beams. As established in previous research on Kalman filter-based methods \citep{lourens2012augmented, aucejo2019practical}, the accuracy of state and input estimation can degrade when measurements are taken from sensors that are relatively far from the location of the input force, or when trying to estimate the response in a location that is poorly observed by the available sensor network. In this specific scenario, when virtual sensing is performed for S7 or S8 (located on the front beam, near the presumed impact zone), these sensors are 'left out'. If the majority of the remaining active sensors are situated at a relative distance from the front beam and the area of impact, the system becomes more susceptible to ill-conditioning \citep{lourens2012augmented}. Beyond this ill-conditioning, the accuracy of virtual sensing can also be influenced by the distribution of the observing sensors across different structural elements. During the leave-one-out process for S7 or S8, no other sensors remain on that same longitudinal beam. If the system connectivity between the longitudinal beams is low, the model may struggle to reconstruct the response accurately \citep{erkaya2018experimental}. This implies that local beam dynamics may become challenging to estimate from distant sensors on parallel members.

\subsection{Optimal real-world sensors placement for virtual sensing}
\label{sec:OSP}
A common question when instrumenting a structure for virtual sensing involves determining the optimal placement of a limited number of sensors to maximize the accuracy of response estimation in the impact direction at a specific, uninstrumented target location. This section outlines a methodology to leverage the developed DT framework to guide an Optimal Sensor Placement (OSP) strategy. The objective is to identify a subset of sensor locations that provides the most informative data for reconstructing the response at a desired target point of interest.

\begin{figure} [H]
    \centering
    \includegraphics[width=1\linewidth]{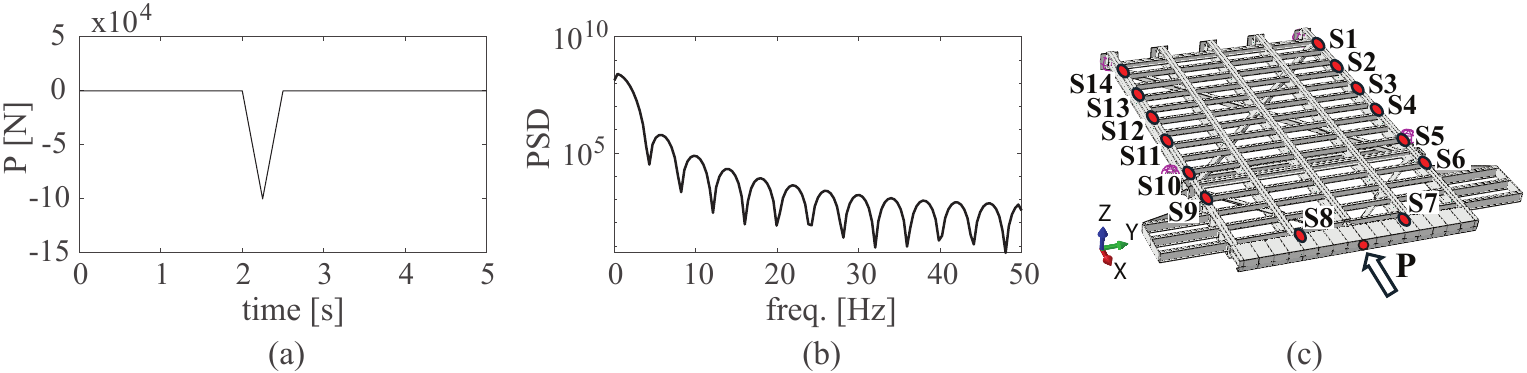}
    \caption{Simulated triangular load used for the OSP numerical study: (a) full time series of the force; (b) power spectral density of the force; (c) sensor and force locations.}
    \label{fig:triangular_load_osp}
\end{figure}

To address this, a numerical simulation was carried out using a sensor selection algorithm. A triangular force, shown in \Cref{fig:triangular_load_osp}, was applied at an arbitrary location on the front beam (\Cref{fig:triangular_load_osp}c) of the FE model to simulate a representative ferry impact. Simulated acceleration responses were generated at fourteen predefined sensor locations, corresponding to those used in Configuration 1 of the experimental campaign (see \Cref{fig_sens_pos}a). The motivation was to ensure comparability with experimental results, allowing validation of the numerical framework against real data. These sensor locations were selected prior to this study based on physical accessibility and to spatially cover the structure as uniformly as possible. No prior OSP strategies had been applied.

\begin{figure}
    \centering
    \includegraphics[width=1\linewidth]{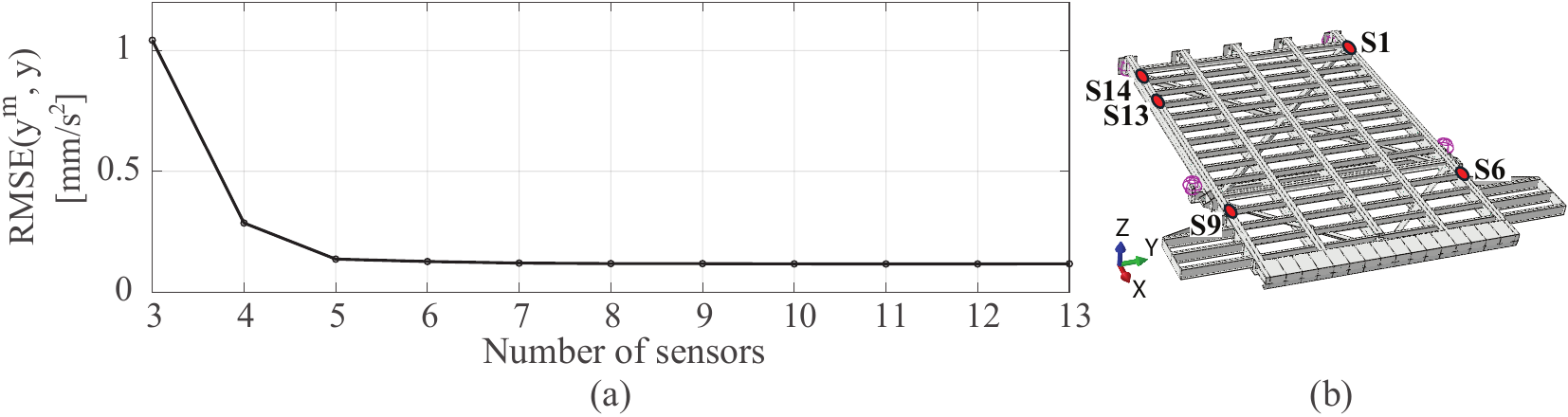}
    \caption{a) Root Mean Squared Error (RMSE) for estimated acceleration response at location S8 as a function of the number of optimally placed sensors; b) five optimal sensor locations.}
    \label{fig:osp_rmse_curve}
\end{figure}

With these simulated responses, a Backward Sequential Sensor Placement (BSSP) strategy \citep{papadimitriou2004optimal} was implemented to identify the optimal sensor subsets for estimating the response at a chosen target location. In this case, sensor location S8 was selected as the target for virtual sensing. The BSSP algorithm works iteratively:
\begin{enumerate}
    \item Initially, all potential sensor locations (excluding the target S8, so thirteen sensors) are considered. The GPLFM is used to estimate the response at S8 using these thirteen simulated sensor inputs, and the RMSE between this estimate and the "true" simulated response at S8 is calculated.
    \item One sensor is removed at a time from the current set. For each possible removal, the GPLFM is re-run with the reduced sensor set to estimate the response at S8, and the corresponding RMSE against the true simulated response is calculated.
    \item The sensor whose removal results in the highest RMSE value is permanently removed from the set. This means the removed sensor was the least critical for estimating the response at S8 given the remaining sensors.
    \item This process is repeated, reducing the number of active sensors by one at each iteration, until a predefined minimum number of sensors is reached, in this case, three sensors.
\end{enumerate}
\Cref{fig:osp_rmse_curve}.a showcases the resulting RMSE values for virtual sensing at location S8 as the number of optimally selected sensors varies from thirteen down to three. As expected, \Cref{fig:osp_rmse_curve}.a illustrates that decreasing the number of sensors leads to an increase in the RMSE, reflecting a loss of spatial information available to the GPLFM for reconstructing the response. Notably, the RMSE remains relatively low and similar for sensor set sizes from thirteen down to approximately five-six sensors. This suggests that beyond five-six optimally placed sensors, adding more sensors yields small returns in terms of improving the estimation accuracy at S8 for this specific simulated load case. However, a more significant increase in RMSE is observed when the number of sensors drops below five, particularly with only three or four sensors, indicating that these smaller sets lack sufficient spatial information to accurately estimate the response at S8.

\begin{figure}
    \centering
    \includegraphics[width=1\linewidth]{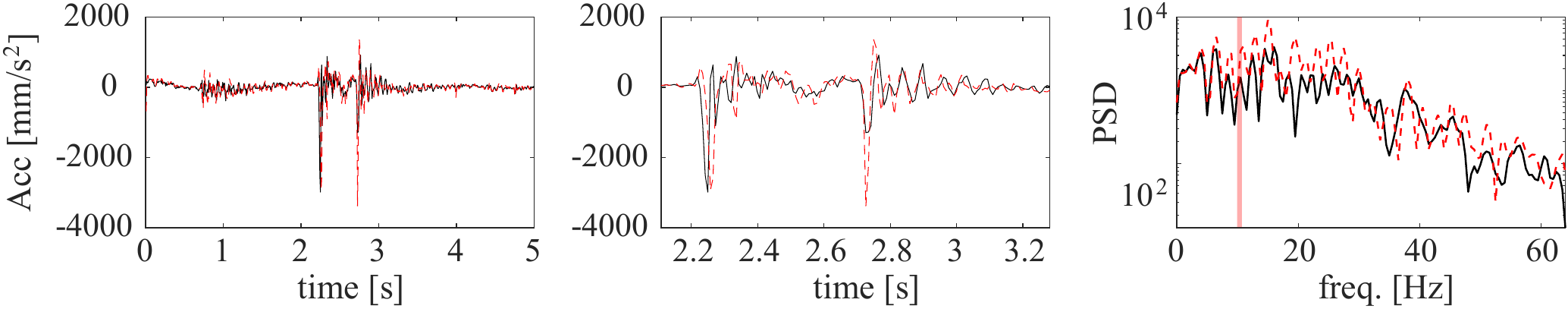}
    \caption{Full time series (left column), zoomed-in view of the time series (center column), and power spectral density (right column) for estimated
(black solid lines) and measured (dashed red line) acceleration responses at location S8. The red vertical bar in the PSD plots indicates the natural frequency of the highest mode included in the reduced-order model.}
    \label{fig:osp_validation_S8}
\end{figure}

Further, the OSP approach is validated with real-world data of configuration 1, the optimal sensor configuration identified for five sensors from the numerical study was selected. These five optimal sensor locations were identified as S1, S6, S9, S13, and S14 (see \Cref{fig:osp_rmse_curve}.b). The GPLFM was then applied using the measured experimental data from these five selected sensors to estimate the acceleration response at location S8. \Cref{fig:osp_validation_S8} shows the estimated response at position S8 against the measured one in the time and frequency domain. The Time Response Assurance Criterion (TRAC), which theoretical background is explained in \ref{app_TRAC}, was used as a metric to evaluate the accuracy of the estimated response in the time domain. The TRAC value between the estimated and measured response at S8 was 3.64\% and it is notably lower than the TRAC of approximately 10\% achieved in the previous section when using thirteen sensors.

Observing the time histories in \Cref{fig:osp_validation_S8}, the estimated acceleration response obtained using the five optimal sensors closely replicates the timing of the peaks seen in the measured response, although the magnitudes differ, similarly to what was observed with thirteen sensors (\Cref{fig:time_freq_S1_S8}). However, there are clear differences in the frequency domain. The PSD of the response estimated with five sensors begins to deviate significantly from the measured PSD after approximately 10 Hz. In contrast, when thirteen sensors were used, this deviation occurred at around 20 Hz. This suggests that while the five optimally chosen sensors capture lower frequency content reasonably well, they may lack the spatial resolution to accurately reconstruct some of the higher frequency modes. 

\subsection{Influence of combining acceleration with displacement data}
\label{sec:lvdt_influence}
Traditional Kalman filter-based state estimation methods are well known to be susceptible to drift when relying solely on acceleration data \citep{azam2015dual}. Accelerometers are insensitive to very low frequencies and quasi-static components. Incorporating displacement data can mitigate drift and improve overall estimation accuracy in such Kalman filter frameworks \citep{azam2015dual}. However, it has also been highlighted that GPLFMs, due to their probabilistic modeling of latent forces and inherent Bayesian inference structure, can often circumvent these drift problems when using only acceleration data \citep{nayek2019gaussian, petersen2022wind}.

For this analysis, sensor configuration 2 was employed, as depicted in \Cref{fig_sens_pos}. This setup comprises five accelerometers (A1-A5) and two LVDTs. The data were originally sampled at 1024 Hz. To facilitate a consistent comparison with analyses performed using sensor configuration 1 (which was sampled at 128 Hz) and to isolate the effect of sensor type rather than sampling rate, the data from configuration 2 were decimated to match the 128 Hz sampling frequency of configuration 1. A detailed investigation into the influence of the sampling frequency is presented later in \Cref{sec:lvdt_influence}.

Virtual sensing was performed using a leave-one-out approach for two target accelerometer locations: A4 and A3 (\Cref{fig_sens_pos}). When estimating the response, the measured data from that specific accelerometer were excluded from the GPLFM and used for validation. Two scenarios were considered for each target: (i) using only the accelerometer data and (ii) using both the accelerometer data and the data from the two LVDTs.

\begin{figure} [H]
    \centering
    \includegraphics[width=1\linewidth]{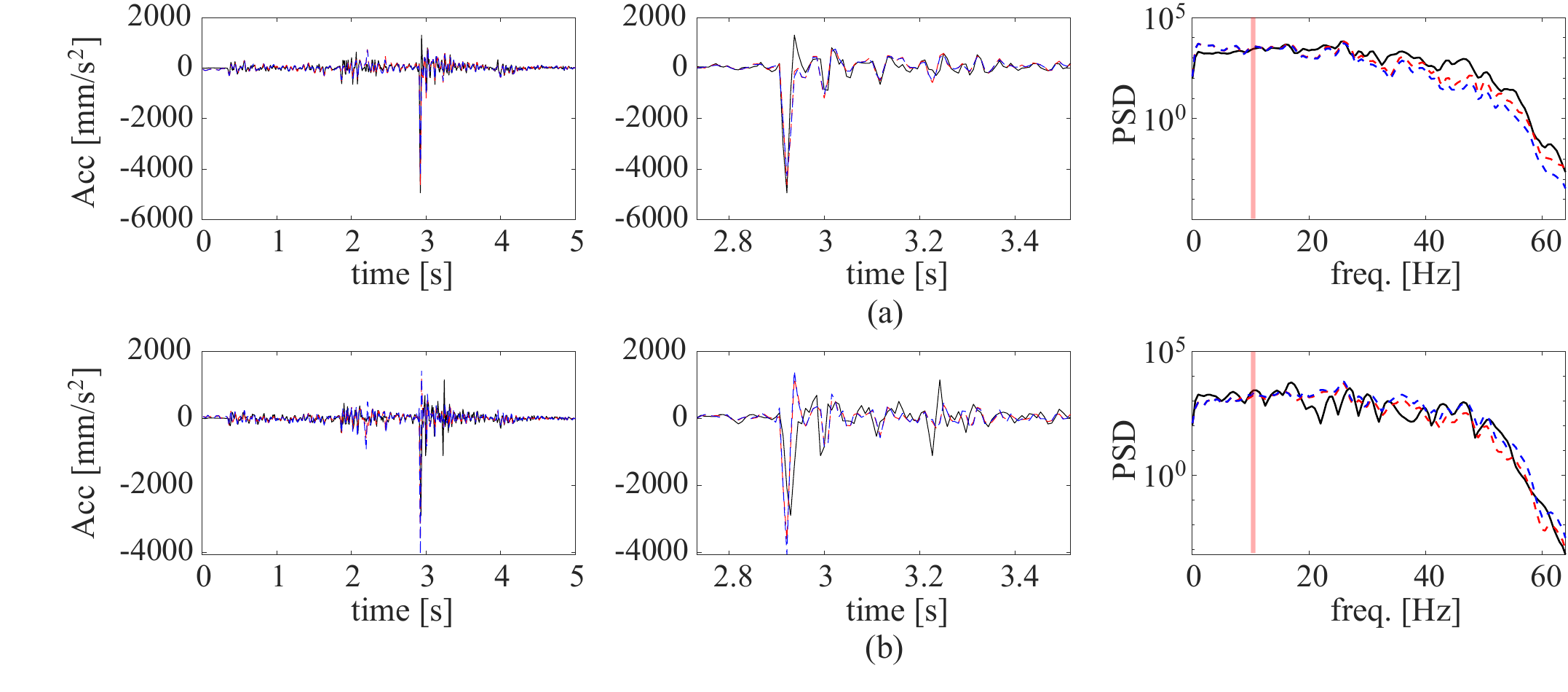}
    \caption{Full time series (left column), zoomed-in view (center column), and power spectral density (right column) of acceleration response sampled at 128 Hz. The results show the estimated response using acceleration data only (dashed red line), and both acceleration and displacement data (dashed blue line), compared with the measured acceleration (solid black line) for (a) sensor location A4; (b) sensor location A3. The red vertical bar in the PSD plots indicates the natural frequency of the highest mode included in the reduced-order model.}
    \label{fig:lvdt_comparison_A4_A3}
\end{figure}

\Cref{fig:lvdt_comparison_A4_A3}a shows the results for location A4, while \Cref{fig:lvdt_comparison_A4_A3}b displays the results for location A3.

The results indicate that adding LVDT data yields very similar acceleration response estimations compared to using acceleration data only, with TRAC values also showing minimal differences between the two scenarios for each location. For location A4, the TRAC value was 75.73\% when using only acceleration data and 74.08\% when including LVDT data. For location A3, the TRAC was 21.72\% for acceleration data only and 16.59\% with the addition of LVDTs. This general similarity further supports the finding that the GPLFM is robust against drift effects even when only accelerometer measurements are available. For the estimation at location A4 (\Cref{fig:lvdt_comparison_A4_A3}a), both scenarios produce high-quality reconstructions in the time domain. Interestingly, in the frequency domain, the estimation using only accelerometers appears to provide a slightly better match to the measured PSD at higher frequencies compared to the case including LVDTs. This very small reduction may be attributed to the displacement data having a mean that is not equal to zero and differing substantially from a Gaussian distribution, which could negatively impact the hyperparameter inference process and lead to a less optimal model. However, it is important to note that the LVDTs would be much more influential if the objective of this study was the estimation of the latent forces and their static component. Consistent with the findings in \Cref{sec:OSP}, the accuracy of the estimation is generally lower for the location situated on the front beam where no other sensors are placed on the same primary longitudinal beam.

\subsection{Sensitivity to an assumed/identified  modal damping ratio}\label{sec:dr_influence}
The analysis employed the GPLFM framework, which, in this implementation, relies on system matrices $\mathbf{A}$ and $\mathbf{B}$ derived from a time-invariant representation of the structure. This assumption implies that the underlying physical properties, such as stiffness, damping, and mass, can be assumed as time-invariant throughout the duration of the dynamic event being modeled. Nonetheless, the identification of damping in complex real-world structures is typically very challenging, and much less accurate than the natural frequency and mode shapes, even when the LTI assumptions hold \citep{Farrar2010}. For this case study, a modal damping ratio of 5\% was assumed for all modes included in the model. 

\begin{figure} [H]
    \centering
    \includegraphics[width=1\linewidth]{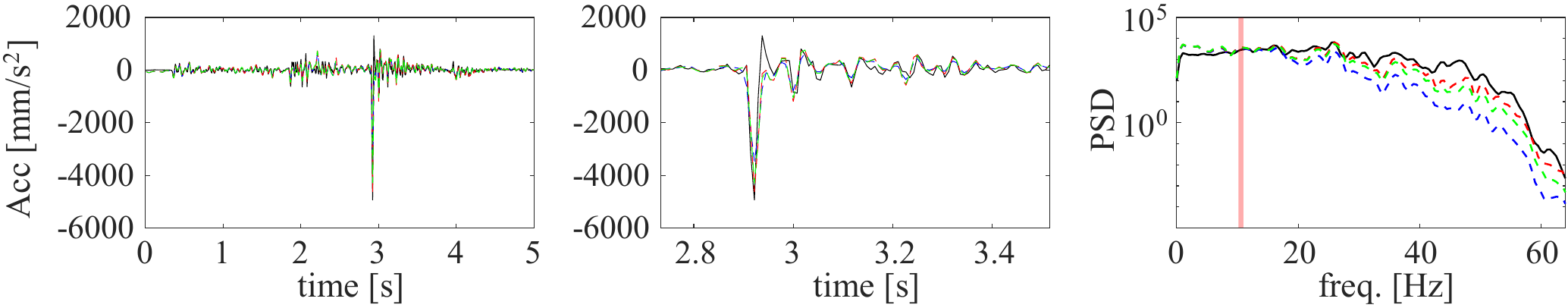}
    \caption{Full time series (left column), zoomed-in view of the time series (center column), and power spectral density (right column) for acceleration response at location A4 with varying modal damping ratio. The results show the estimated response assuming a modal damping ratio of 2\% (dashed blue line), a modal damping ratio of 5\% (dashed red line), and a modal damping ratio of 8\% (dashed green line) compared with the measured acceleration (solid black line). The red vertical bar in the PSD plots indicates the natural frequency of the highest mode included in the reduced-order model.}
    \label{fig:damping_sensitivity}
\end{figure}

To evaluate the sensitivity of the GPLFM to potential inaccuracies in this assumed damping ratio, two perturbation scenarios were investigated using sensor configuration 2 and measurement data decimated to 128 Hz: one underestimating the damping ratio at 2\% and another overestimating it at 8\%. The resulting estimated acceleration responses at position A4 for the baseline (5\%) and perturbed damping models (2\% and 8\%) are compared against the measured data in \Cref{fig:damping_sensitivity}.

Observation of the time domain responses in \Cref{fig:damping_sensitivity} shows that the estimated accelerations are similar across all three assumed damping ratios (2\%, 5\%, and 8\%). All models accurately reconstruct the overall shape of the measured response, and specifically, the timing and approximate amplitude of the main acceleration peak corresponding to the ferry impact are well captured. This suggests a high degree of robustness in the time-domain estimation to the assumed level of modal damping. The TRAC values are 75.73\% for the 5\% damping, 74.22\% for 8\% damping, and a slightly lower 66.77\% for 2\% damping. While the 2\% damping scenario shows a reduction in TRAC, all values indicate a good level of correlation and suggest a high degree of robustness in the time-domain estimation to the assumed level of modal damping. In the frequency domain, the PSD of the estimated responses for all three damping scenarios closely matches the measured PSD up to approximately 20 Hz. The estimation with a 2\% damping ratio tends to exhibit a slightly larger error with respect to the measured PSD at these higher frequencies compared to the 5\% and 8\% cases. This suggests that while the GPLFM can effectively compensate for damping inaccuracies at lower frequencies, its ability to do so may decrease at higher frequencies, especially when the assumed damping is significantly lower than the true (unknown) system damping. 

\subsection{Sensitivity to sampling frequency}\label{sec:lvdt_influence}
\begin{figure} [H]
    \centering
    \includegraphics[width=1\linewidth]{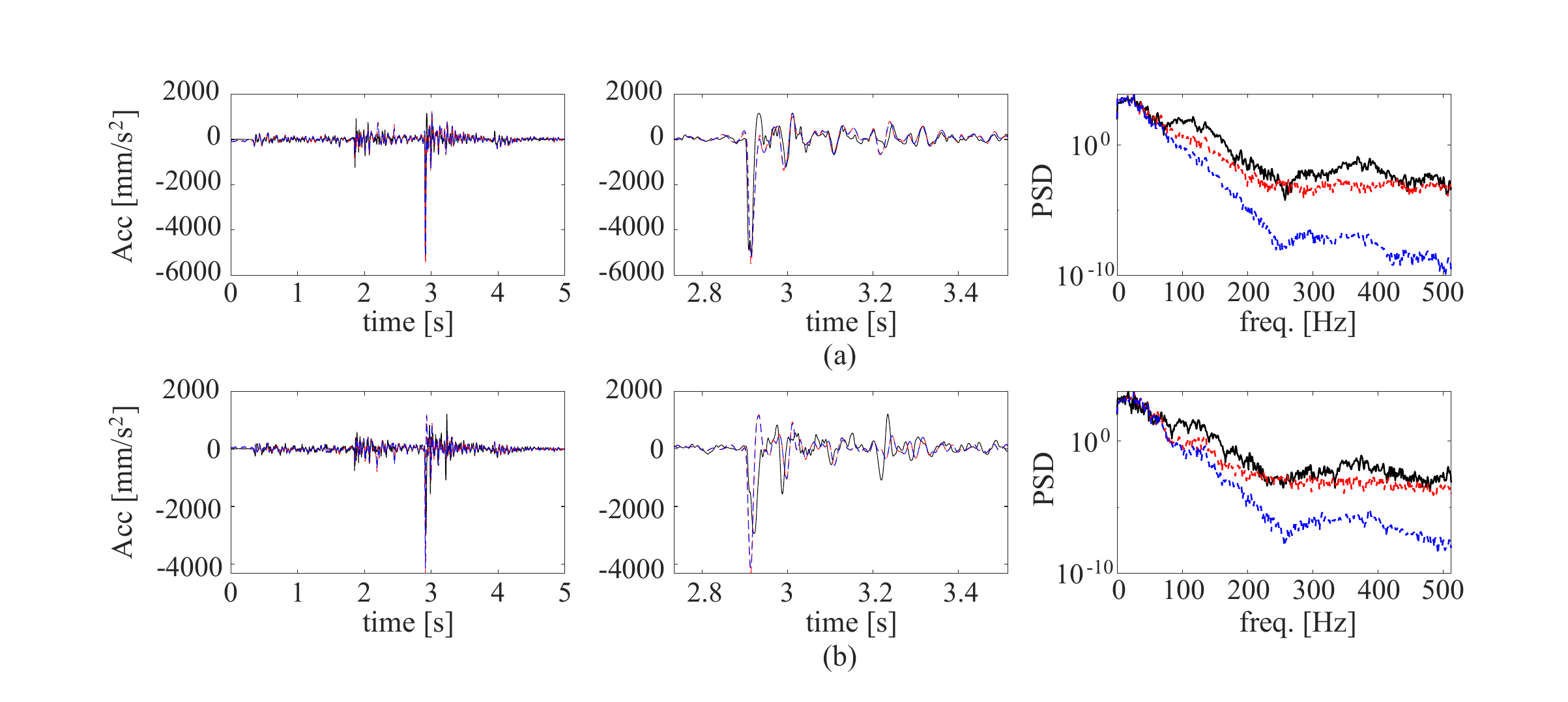}
    \caption{Full time series (left column), zoomed-in view (center column), and power spectral density (right column) of the acceleration response at location S13 sampled at 1024 Hz. The results show the estimated response using acceleration data only (dashed red line), and both acceleration and displacement data (dashed blue line), compared with the measured acceleration (solid black line).}
    \label{fig:response_1024Hz_A4}
\end{figure}

The choice of sampling frequency in data acquisition directly impacts the ability to capture the full spectrum of a dynamic event.  This is particularly important when dealing with impact events of short durations, which would excite high frequencies. Higher sampling rates can capture transients and higher-frequency components that might be aliased or lost at lower rates. This section investigates the performance of the GPLFM when utilizing high sampling frequency data and compares estimations made with and without LVDT data. For this analysis, sensor configuration 2 was employed, with data recorded at its original sampling frequency of 1024 Hz. The virtual sensing estimation was performed for accelerometer location A4, with its own data excluded from the GPLFM input and used for validation. The estimated and measured responses are shown in \Cref{fig:response_1024Hz_A4}. 

In the time domain, the estimated responses using acceleration data only and with LVDTs provide good and similar reconstructions of the measured acceleration at A4. The TRAC value was 69.71\% when using only acceleration data, and slightly lower at 68.14\% when LVDT data was also included. These values are slightly lower than those obtained for the same location A4 when using data decimated to 128 Hz (which were 75.73\% for accelerometers only and 74.08\% with LVDTs. This performance is likely attributable to the fact that the dominant dynamic response of this ferry impact occurs within the lower frequency range (below 50 Hz). As shown previously, the GPLFM matched the measured response within these lower frequencies. Observing the PSD of the measured response (solid black line), it is evident that the majority of the signal's energy content is concentrated below approximately 40-50 Hz for this impact and sensor configuration. Beyond 150 Hz, the energy in the measured signal decreases substantially.

Interestingly, a notable difference between the responses estimated using acceleration data only and with LVDTs emerges at higher frequencies. The PSD estimated using accelerometers and LVDTs (dashed blue line) exhibits a significant decrease in energy at frequencies above approximately 100 Hz compared to the estimation using accelerometers only (dashed red line). This deviation may be due to the LVDTs providing no useful information in this higher frequency range. When their data is included in the GPLFM process, it might influence the hyperparameter optimization or the noise covariance estimations in a way that leads the model to attenuate the overall high-frequency content of the joint estimation more than when only accelerometers are used.

\section{Conclusions}\label{sec:conclusions}
This paper investigates the development of a DT of a ferry quay by combining a FE model with real-world measurements collected at selected locations during ferry impacts, in order to estimate virtual responses across the entire structure. This is carried out by using a PEML strategy based on a GPLFM. The DT integrates a physics-based FE model of the linkspan (reduced to its first seven modes) with a data-driven GP to model the unknown ferry impact forces through their modal components. The analyses presented were based on data collected during two distinct ferry impact events, utilizing different sensor configurations and sampling frequencies.

A significant challenge in modeling this type of operational infrastructure lies in its time-varying behavior, primarily driven by the interaction with the ferry and the influence of environmental conditions. To address this complexity, the present study assessed the virtual sensing capabilities of the GPLFM specifically during the ferry impact phase, under the simplifying assumption that the system behaves as LTI during this short time window. Furthermore, unknown impact forces were modeled through their modal components, overcoming the need to specify the location and direction of the ferry impact, which were treated as unknown. 

An investigation into the influence of sensor location, conducted using a leave-one-out cross-validation with a sensor configuration comprising fourteen accelerometers, demonstrated that the GPLFM generally provided accurate acceleration estimations at most locations. To assess the robustness of the GPLFM predictions, different hybrid physics–data model configurations were explored. It was found that adding LVDT measurements to the acceleration measurements yielded very similar acceleration estimation accuracy to using accelerometers alone, aligning with previous findings conducted on Kalman filter methods based on Latent Force Models \citep{petersen2022wind}. When investigating the influence of incorrectly assumed modal damping ratio (tested at 2\%, 5\%, and 8\%), time-domain estimations remained consistent, with the 2\% damping case showing more deviation at higher frequencies. The influence of the sampling frequency was also investigated, showing that virtual sensing results obtained at 128 Hz were comparable to those at 1024 Hz. A numerical study was performed to support optimal sensor placement to improve virtual sensing accuracy using a BSSP strategy to identify the most informative sensor locations. The results indicated that a configuration with only five strategically placed sensors could provide a response estimation with slightly reduced, but still comparable, accuracy to that obtained with the full set of fourteen sensors. 

In conclusion, this study assessed the applicability of the GPLFM for virtual sensing of accelerations on an operational ferry quay characterized by complex boundary conditions, time-invariant behavior during the analyzed window, and unknown impact loads. The method proved capable of providing useful response estimates even with sparse sensor data and under strong modeling assumptions. However, it was found that the performance decreased when estimating responses at locations on the front beam. These sensors were (i) near the impact zone, and (ii) located on longitudinal beams where no other sensors were placed. While this lower accuracy may be related to modeling errors, such as the LTI simplification, it also aligns with known challenges in state estimation: (1) ill-conditioning, common in KF-based methods when observation points are distant from the excitation/estimation zone \citep{lourens2012augmented, maes2016verification}, and (2) the difficulty of estimating responses across insufficiently stiff connections between parallel members \citep{erkaya2018experimental}. Future research should address the following challenges to enhance the capabilities of the Digital Twin further. A primary direction involves extending the GPLFM to incorporate a switching model that accounts for the time-dependency of system matrices and damping, particularly during transitions between disconnected and docked states, to enable a complete description of the DT. Another significant improvement is the inclusion of nonlinear parameters, such as the stiffness of the fenders and lifting towers, to better represent the real-world behavior of the structure. Additionally, modeling the damping ratio as a latent force governed by a GP may further enhance the algorithm's ability to adapt to dynamic changes in system behavior.

\begin{appendix}
\section{Root Mean Square Error and Time Response Assurance Criterion mathematical background}\label{app_TRAC}
To assess the similarity of two time-domain signals, two error metrics are considered in this paper: the Root Mean Square Error (RMSE) and the Time Response Assurance Criterion (TRAC). These metrics quantify the correlation and magnitude differences.

The RMSE quantifies the root of the average squared deviation between the two signals, offering a measure of the estimation error magnitude. It is defined as:
\begin{equation}
    RMSE (\varepsilon, \hat{\varepsilon}) = \sqrt{\frac{1}{N} \sum_{k=1}^{N} \left(\hat{\varepsilon}[k] - \varepsilon[k]\right)^2}
\end{equation}
where $\left(\hat{\varepsilon}[k] - \varepsilon[k]\right)^2$ is the squared difference at each time step $k$.

On the other hand, the TRAC metric quantifies the correlation between two vectors, $\hat{\varepsilon}$ and $\varepsilon$, and is defined as:
\begin{equation}
    TRAC (\varepsilon, \hat{\varepsilon}) = \frac{(\hat{\varepsilon} \varepsilon^T)^2}{(\hat{\varepsilon} \hat{\varepsilon}^T)(\varepsilon \varepsilon^T)}
\end{equation}
where $\varepsilon \in \mathbb{R}^N$ and $\hat{\varepsilon} \in \mathbb{R}^N$ represent the compared signals, and $N$ is the number of time samples. TRAC values range from 0\%, indicating no correlation, to 100\%, signifying a perfect match.

\section{Matérn 3/2 covariance function state-space model}\label{app:ssm}
The Matérn kernel defines the covariance between two points $t$ and $t'$ and is generally expressed as: 
\begin{equation}\label{matern}
\kappa(r; \nu, \alpha, l) = \alpha^2 \frac{2^{1-\nu}}{\Gamma(\nu)} \left( \frac{\sqrt{2 \nu} r}{l} \right)^\nu K_\nu \left( \frac{\sqrt{2 \nu} r}{l} \right)
\end{equation}
where $r = \lVert t - t' \rVert$ is the Euclidean distance between the two points $t$ and $t'$. $\alpha^2$, $l$, and $\nu$ are positive hyperparameters that define the behavior of the Gaussian process (GP). Specifically, $\alpha^2$ influences the overall variance, $l$ controls the length scale and how quickly the correlation between points in the GP decreases with distance, while the hyperparameter $\nu$ defines the smoothness of the GP model and specifies the number of derivatives that the GP is differentiable. The term $\Gamma(\nu)$ is the Gamma function, and $K_\nu$ is the modified Bessel function of the second kind. 

GPs with a Matérn covariance function may be formulated into a state-space representation under the assumption that the spectral density of the kernel has a rational form. The derivation follows a sequence of mathematical steps, which are briefly outlined below. For a more detailed derivation, the reader is referred to \cite{vettori2024assessment}.

The spectral density of Matérn family covariance functions is derived by the Wiener-Khinchin theorem \citep{chatfield2019analysis} and expressed as:
\begin{equation}\label{maternsd}
S(\omega) = \alpha^2 \frac{2 \pi^{1/2} \Gamma(\nu + 1/2)}{\Gamma(\nu)} \lambda^{2\nu} (\lambda^2 + \omega^2)^{-(\nu - 1/2)}.
\end{equation}
By setting $\lambda = \sqrt{2\nu}/\ell$ and expressing $\nu$ in terms of an integer $p$ as $\nu = \mu + 1/2$, \cref{maternsd} simplifies to:
\begin{equation}\label{maternsspro}
S(\omega) \propto (\lambda^2 + \omega^2)^{-(\mu+1)}.
\end{equation}
To express $S(\omega)$ in a form suitable for an SSM and allow the derivation of a stable transfer function representation, \cref{maternsspro} is spectrally factorized as: 
\begin{equation}\label{maternssfact}
S(\omega) \propto (\lambda + i\omega)^{-(\mu-1)} (\lambda - i\omega)^{-(\mu+1)}.
\end{equation}
which facilitates its interpretation as the spectral density of the output of a system $F(\omega)$ with transfer function $H(i\omega)$ and excited by an input $w(t)$ with a spectral density $q_c$. The term $F(\omega)$ corresponds to the Fourier transform of $f(t)$ formulated as:
\begin{equation}\label{maternsystemout}
F(\omega) = H(i\omega)W(\omega).
\end{equation}
where $W(\omega)$ is the Fourier transform of the excitation term $w(t)$.

This transfer function $H(i\omega)$ can be directly converted into a continuous-time state-space representation \citep{hartikainen2010kalman}. The state-space matrices $\mathbf{F}, \mathbf{L} \text{ and } \mathbf{H}$ characterize the evolution of the latent process in a Markovian framework. The state-space matrices associated to the Matérn 3/2 kernel, which corresponds to the case $\mu = 1$, i.e., $\nu = 3/2$, are defined as:
\begin{equation}
\mathbf{F}_c =
\begin{bmatrix}
0 & 1 \\
-\lambda^2 & -2\lambda
\end{bmatrix}, \quad
\mathbf{L}_c = \begin{bmatrix} 0 \\ 1 \end{bmatrix}, \quad
\mathbf{H}_c = \begin{bmatrix} 1 & 0 \end{bmatrix}.
\end{equation}

\section{Mode shapes used to build the reduced-order model}\label{app:mode_shapes}
\begin{figure} [H]
    \centering
\includegraphics[width=0.7\linewidth]{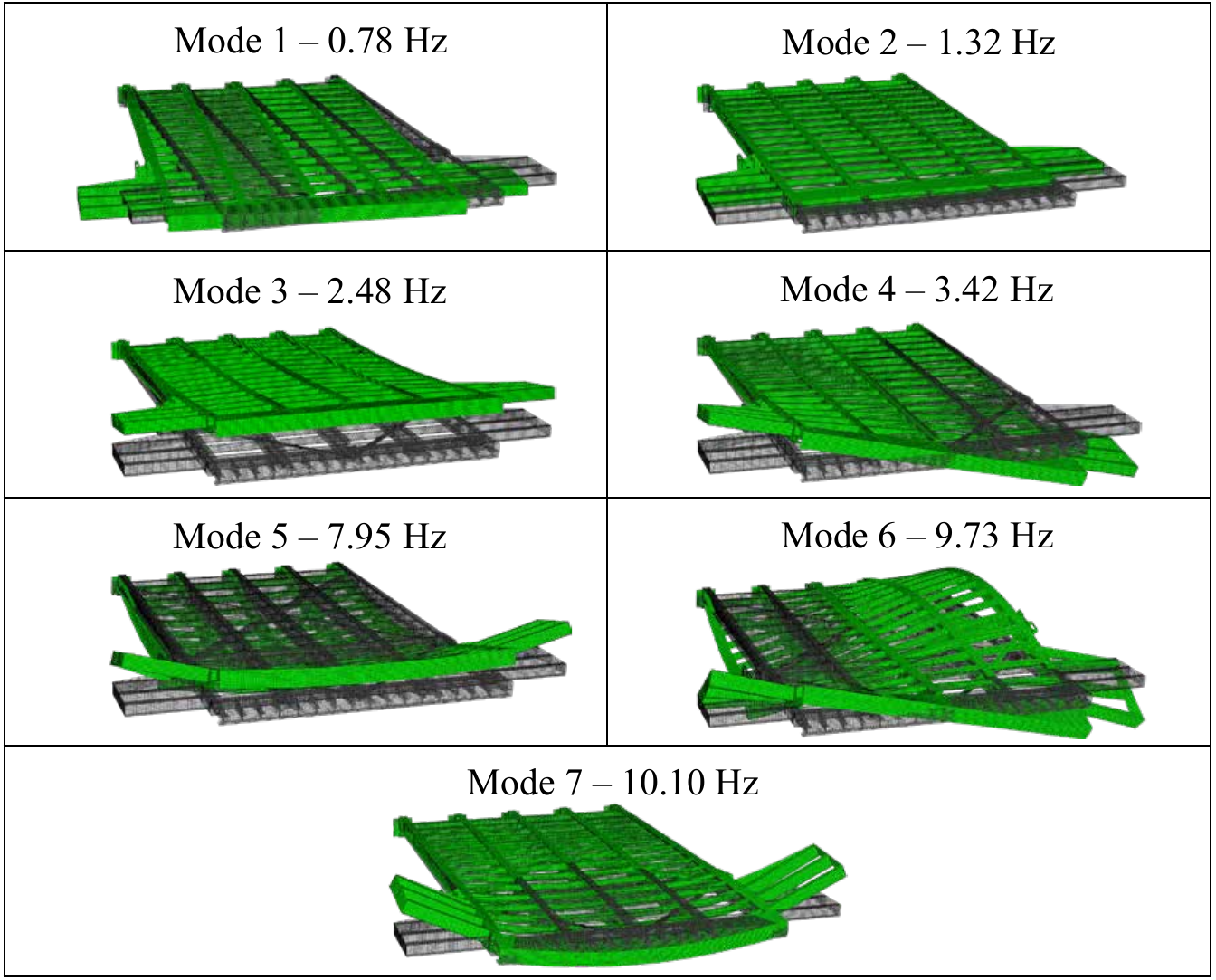}
    \caption{Modal shapes of the FE model of the Magerholm ferry quay}
    \label{fig:enter-label}
\end{figure}
\end{appendix}

\begin{Backmatter}

\paragraph{Acknowledgments}
The authors wish to acknowledge the support from Møre og Romsdal Fylke for providing operational and financial support for the installation and maintenance of the sensor system. The authors are grateful for the extensive help from Dr. Ingrid Anne Lervik from the Møre og Romsdal County Municipality.

\paragraph{Funding Statement}
This research was supported by grants from the Research Council of Norway through the SARTORIUS project (Project No.~353029).

\paragraph{Competing Interests}
The authors declare no competing interests exist.

\paragraph{Ethical Standards}
This research complies with all applicable ethical standards and legal requirements of the country in which the study was conducted. All data acquisition procedures were carried out in accordance with institutional and national regulations, ensuring safety, transparency, and responsible conduct of research.

\paragraph{Author Contributions}
Conceptualization: All authors; Methodology: L.S., A.C.; Investigation: all authors; Data curation: L.S.; Data Visualization: L.S.; Writing original draft: L.S.; Writing – review \& editing: all authors.; Supervision: T.N., A.C.; Project management: T.N.; All authors approved the final submitted draft.

\clearpage


\printbibliography

\end{Backmatter}

\end{document}